\documentclass[english,11pt,a4paper]{article}
\usepackage[T1]{fontenc}
\usepackage[latin9]{inputenc}
\usepackage{amsmath}
\usepackage{amsthm}
\usepackage{amssymb}
\usepackage{graphicx}

\makeatletter

\providecommand{\tabularnewline}{\\}

\numberwithin{equation}{section}

\usepackage{jheppub}
\usepackage{orcidlink}
\allowdisplaybreaks

\usepackage{makecell} 
\usepackage{simpler-wick}

\makeatother

\usepackage{babel}
\begin{document}
\title{Dark Photons in the Early Universe: From Thermal Production
to Cosmological Constraints}

\abstract{ Dark photons, a generic class of light gauge bosons that
interact with the Standard Model (SM) exclusively through kinetic
mixing, arise naturally in many gauge extensions of the SM. Motivated
by these theoretical considerations, we present a comprehensive analysis
of their thermal production in the early universe. Our calculation
covers a broad range of dark photon masses from 0.1  keV to 100 MeV
and include inverse decay, annihilation, and semi-Compton processes.
Wherever possible, we present analytical estimates of the production
rates and yields, and verify their accuracy numerically. For dark
photons lighter than twice the electron masses (around 1 MeV), we
find that our analytical estimate of the freeze-in yield based on
resonant production is very accurate,  implying that off-resonance
contributions can be neglected in practice. For heavy dark photons,
although this conclusion no longer holds, we derive an interesting
ratio, $4\pi e/27\approx0.14$, with $e$ the coupling constant of
QED, that can be used to estimate the relative importance of on- and
off-resonance contributions. Finally, using the calculated abundance
of dark photons in the early universe, we derive cosmological constraints
on the dark photon mass and kinetic mixing. Compared with bounds from
stellar cooling and supernovae, the cosmological constraints are most
stringent in the mass range from 0.1 MeV to 6 MeV, within which kinetic
mixing at the level of $10^{-12}\sim10^{-10}$ can be probed.

}

\author[a]{Xun-Jie Xu \orcidlink{0000-0003-3181-1386}}
\author[a,b]{and Boting Zhou \orcidlink{0009-0007-4432-5623}}
\affiliation[a]{Institute of High Energy Physics, Chinese Academy of Sciences, Beijing 100049, China}
\affiliation[b]{School of Physical Sciences, University of Chinese Academy of Sciences, Beijing 100049, China}
\emailAdd{xuxj@ihep.ac.cn} 
\emailAdd{zhoubt@ihep.ac.cn} 
\preprint{\today}  
\maketitle

\section{Introduction}

Among a plethora of extensions of the Standard Model (SM), one of
the most extensively studied cases is gauge extensions. The simplest
gauge extension, introducing an extra gauged $U(1)$, has an interesting
feature: the associated gauge boson  can couple through a kinetic
mixing term to the hypercharge gauge boson of the SM~\cite{Holdom:1985ag}.
When the new gauge boson is light (well below the electroweak scale),
the kinetic mixing gives rise to photon-like interactions with SM
particles; that is, its effective couplings to SM particles are proportional
to their electric charges, with a universal suppression set by the
kinetic-mixing parameter. Widely referred to as the dark photon,
this light, weakly-coupled gauge boson has drawn considerable attention
due to its rich phenomenology---see Refs.~\cite{Jaeckel:2010ni,Essig:2013lka,Alexander:2016aln,Ilten:2018crw,Bauer:2018onh,Fabbrichesi:2020wbt,Caputo:2021eaa}
for recent reviews---including potential connections with the dark
matter problem of the SM~\cite{Redondo:2008ec,McDermott:2019lch,Cyncynates:2024yxm,Trost:2024ciu,Hook:2025pbn,Zhang:2025pgb,Vogl:2024ack},
and has also motivated numerous laboratory searches~\cite{Belle-II:2022jyy,Feng:2017uoz,FASER:2023zcr,Alekhin:2015byh,Gardner:2015wea,Berlin:2018pwi,Chou:2016lxi,MATHUSLA:2020uve}.

Complementary to laboratory searches, cosmology provides a powerful
avenue for probing dark photons through their observable imprints
on the  evolution of the early universe~\cite{Redondo:2008ec,Fradette:2014sza,Berger:2016vxi,Pospelov:2018kdh,Ibe:2019gpv,McDermott:2019lch,Coffey:2020oir,Li:2020roy,Caputo:2022keo,Adshead:2022ovo,Pirvu:2023lch,Gan:2023wnp,Cyncynates:2023zwj,Aramburo-Garcia:2024cbz,McCarthy:2024ozh,Trost:2024ciu,Cyncynates:2024yxm,Hook:2025pbn,Jaeckel:2008fi}.
Depending on whether the mass is heavy (above the MeV scale)~\cite{Berger:2016vxi,Ibe:2019gpv,Coffey:2020oir,Li:2020roy,Adshead:2022ovo},
light (from keV to MeV)~\cite{Redondo:2008ec,Fradette:2014sza,Gan:2023wnp},
or ultra-light (well below the aforementioned scales)~\cite{Pospelov:2018kdh,McDermott:2019lch,Caputo:2022keo,Pirvu:2023lch,Cyncynates:2023zwj,Aramburo-Garcia:2024cbz,McCarthy:2024ozh,Trost:2024ciu,Cyncynates:2024yxm,Hook:2025pbn,Jaeckel:2008fi},
dark photons can be produced in multiple ways in the early universe.
Heavy dark photons with kinematically-allowed decay channels to charged
fermions can be efficiently produced via inverse decays from charged
fermions in the thermal bath. Light dark photons are produced via
scattering processes, which in certain regimes can be approximated
by resonant conversion (or oscillation) between photons and dark photons.
Ultra-light dark photons are often assumed to be produced non-thermally,
e.g., via the misalignment mechanism.  

In this work, we focus on the thermal production of dark photons in
the early universe, which is essentially unavoidable in the thermal
environment if the kinetic mixing is sizable.   Although  this
has been computed in a few studies for various purposes (see, e.g.,
\cite{Redondo:2008ec,Ibe:2019gpv,Li:2020roy}), a consistent treatment
applicable to both the heavy and light mass regimes is still lacking.
Moreover, existing calculations require solving the Boltzmann equation
numerically,  whereas we attempt to push the analytical approach
as far as possible. We show that, for both inverse decay and scattering
processes, analytical collision terms can be obtained with sufficient
accuracy for quantitative studies. In addition, the dark-photon yield
 in the freeze-in regime can also be estimated analytically. In particular,
we  derive an interesting ratio of resonant to post-resonance yields
for heavy dark photons: $4\pi e/27\approx0.14$, where $e$ is the
QED coupling. This implies that, for heavy dark photons, resonant
production provides a significant, though subdominant, contribution.
  We also apply our calculation to derive cosmological constraints
on dark photons across a broad mass range, from 0.1 keV to 10 MeV.
The constraints are derived with rather robust and conservative energy-density
considerations, and we show that they are complementary to stellar
cooling and supernovae bounds. 

The structure of this paper is organized as follows. In Sec.~\ref{sec:formulation},
we introduce the basic formalism including the Lagrangian for the
dark photon, its interactions, and the Boltzmann equation. In Sec.\ \ref{sec:Thermal-processes},
we calculate the collision terms of relevant thermal processes, without
including the medium effect (also known as the plasma effect), which
is elaborated in Sec.\ \ref{sec:medium-effect}. In Sec.\ \ref{sec:Cosmological-evolution},
we discuss the evolution of dark photons and compare analytical results
obtained in the previous sections with numerical solutions of the
Boltzmann equation. Sec.\ \ref{sec:Cosmological-constraints} presents
the cosmological constraints derived in this work.  Finally, we conclude
in Sec.~\ref{sec:conclusion} and relegate some details to the appendices.

\section{Lagrangian and Boltzmann equation \label{sec:formulation}}

In a gauged $U(1)$ extension of the SM, the gauge field associated
with the extra $U(1)$ can couple  to the gauge field of $U(1)_{Y}$
in the SM via the following kinetic mixing term~\cite{Holdom:1985ag}:
\begin{equation}
{\cal L}\supset-\frac{\epsilon}{2}F^{\mu\nu}F'_{\mu\nu}\thinspace,\label{eq:FF}
\end{equation}
where $\epsilon$ is a coupling constant; and $F^{\mu\nu}$ and $F'_{\mu\nu}$
denote the field strength tensors of $U(1)_{Y}$ and the extra $U(1)$,
respectively. After canonicalizing the kinetic terms and diagonalizing
the mass terms of all gauge bosons (see, e.g.,  \cite{Lindner:2018kjo,Li:2023vpv}),
one obtains gauge fields in mass eigenstates with properly normalized
kinetic terms. Among the mass eigenstates, one should be massless
due to the unbroken electromagnetic $U(1)$, and another can be relatively
light compared with  the electroweak scale. The former is identified
as the photon (with the field denoted by $A_{\mu}$ and the particle
by $\gamma$) whereas the latter is defined as the dark photon (with
the field denoted by $A'_{\mu}$ and the particle by $\gamma'$).
 The mass of the dark photon, denoted by $m_{\gamma'}$, may arise
from spontaneous symmetry breaking of the extra $U(1)$~\cite{Berger:2016vxi,Lindner:2018kjo}
or the St\"uckelberg mechanism~\cite{Stueckelberg:1938hvi,Feldman:2007wj}. 

The canonicalization of kinetic terms gives rise to interactions of
$A$ and $A'$ with SM fermions:
\begin{equation}
{\cal L}\supset eJ_{{\rm em}}^{\mu}\left(A_{\mu}+\varepsilon A'_{\mu}\right),\label{eq:J-A-A}
\end{equation}
where $e=\sqrt{4\pi\alpha}$ with $\alpha\approx1/137$ the fine structure
constant, $J_{{\rm em}}^{\mu}$ is the electric current, and $\varepsilon=\epsilon\cos\theta_{W}$
with $\theta_{W}$ the Weinberg angle. We note here that in the basis
of mass eigenstates, the kinetic mixing has been removed and $A'$
interacts with the SM only via Eq.~\eqref{eq:J-A-A}. 

In the early universe, the dark photon can be produced through particle
scattering processes in the thermal bath, and can also be absorbed
through the corresponding inverse processes. Its phase space distribution
function, denoted by $f_{\gamma'}(t,p)$, obey the Boltzmann equation:
\begin{equation}
\left[\frac{\partial}{\partial t}-Hp\frac{\partial}{\partial p}\right]f_{\gamma'}(t,p)=\Gamma_{\gamma'}^{{\rm gain}}\left(1+f_{\gamma'}\right)-\Gamma_{\gamma'}^{{\rm loss}}f_{\gamma'},\label{eq:dfdt}
\end{equation}
where $H=\dot{a}/a$ is the Hubble parameter with $a$ the scale factor,
and $\Gamma_{\gamma'}^{{\rm gain/loss}}$ denotes the gain or loss
rate of $\gamma'$ due to dark photon interactions. In the radiation-dominated
era, the Hubble parameter is given by
\begin{equation}
H=g_{H}\frac{T^{2}}{m_{{\rm pl}}}\thinspace,\text{with}\ \ g_{H}\equiv1.66g_{\star}^{1/2}\thinspace.\label{eq:H}
\end{equation}
Here, $T$ is the temperature of the thermal bath, $m_{{\rm pl}}\approx1.22\times10^{19}$
GeV is the Planck mass, and $g_{\star}$ denotes the effective number
of relativistic degrees of freedom. 

When $f_{\gamma'}\ll1$, corresponding to the freeze-in regime, one
may neglect the backreaction term $\Gamma_{\gamma'}^{{\rm loss}}f_{\gamma'}$
in Eq.~\eqref{eq:dfdt} and write it in the following integral form:
\begin{equation}
f_{\gamma'}(a,\ p)\approx\int_{0}^{a}\frac{\Gamma_{\gamma'}^{{\rm gain}}(a',\ ap/a')}{H(a')a'}da'\thinspace.\label{eq:f-int}
\end{equation}
Here,   $f_{\gamma'}$ and $\Gamma_{\gamma'}^{{\rm gain}}$ have
been re-defined as functions of $a$ and $p$, i.e., $f_{\gamma'}=f_{\gamma'}(a,\ p)$
and $\Gamma_{\gamma'}^{{\rm gain}}=\Gamma_{\gamma'}^{{\rm gain}}(a,\ p)$.
Equation~\eqref{eq:f-int} is derived through a change of variables
from $t$ to $a$ and $p$ to the comoving momentum---see, e.g.,
 Appendix B of Ref.~\cite{Li:2022bpp} for a detailed derivation.

In this work, we also consider the number density of $\gamma'$, denoted
$n_{\gamma'}$, which satisfies the following Boltzmann equation:
\begin{equation}
\frac{dn_{\gamma'}}{dt}+3Hn_{\gamma'}=C_{\gamma'}^{{\rm gain}}-C_{\gamma'}^{{\rm loss}}\thinspace,\label{eq:dndt}
\end{equation}
where the collision terms $C_{\gamma'}^{{\rm gain}}$ and $C_{\gamma'}^{{\rm loss}}$
are the phase space integrals of the gain and loss terms in Eq.~\eqref{eq:dfdt},
respectively. Similar to Eq.~\eqref{eq:f-int}, one can derive an
integral form for $n_{\gamma'}$ in the freeze-in regime:
\begin{equation}
n_{\gamma'}a^{3}\approx\int_{0}^{a}\frac{a'{}^{2}C_{\gamma'}^{{\rm gain}}(a')}{H(a')}da'\thinspace,\label{eq:n-int}
\end{equation}
with 
\begin{equation}
C_{\gamma'}^{{\rm gain}}\approx\int g_{\gamma'}\frac{d^{3}p}{(2\pi)^{3}}\Gamma_{\gamma'}^{{\rm gain}}\thinspace.\label{eq:Gamma-to-C}
\end{equation}
Here $g_{\gamma'}$ denotes the number of degrees of freedom of $\gamma'$. 

Throughout this paper, we often divide $n_{\gamma'}$ by the entropy
density 
\begin{equation}
s=g_{\star,s}\frac{2\pi^{2}T^{3}}{45}\thinspace,\label{eq:-61}
\end{equation}
where $g_{\star,s}$ is the effective number of relativistic degrees
of freedom in entropy. In the absence of interactions (or when production
and deletion processes are inactive), $n_{\gamma'}/s$ is approximately
constant. 

\section{Thermal processes\label{sec:Thermal-processes}}

We consider dark photon thermal production in the early universe when
it is dominated by photons ($\gamma$), electrons ($e^{\pm}$) and
neutrinos ($\nu$/$\overline{\nu}$), corresponding to an epoch of
$1\ \text{eV}\lesssim T\lesssim100\ \text{MeV}$. Below $\sim1\ \text{eV}$,
the universe undergoes matter-radiation equality ($0.8$ eV) and recombination
($0.3$ eV), which are not included in our analysis.   Above $\sim100$
MeV, one needs to include more thermal species ($\mu^{\pm}$, $\pi^{\pm}$,
etc.) in the analysis while the corresponding yield of $\gamma'$
produced at such temperatures is subdominant for light $\gamma'$.
Hence for simplicity, we neglect this part. 

\begin{figure}
\centering

\includegraphics[width=0.95\textwidth]{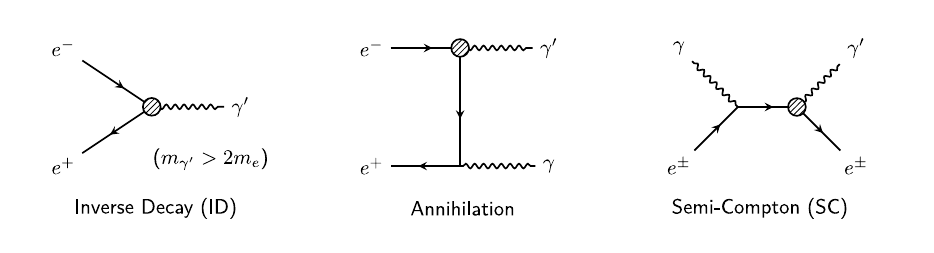}

\caption{\label{fig:feyn} Thermal processes for dark photon production, including
inverse decay (left), annihilation (middle), and semi-Compton (right).
 The hatched blobs indicate the effective coupling of $\gamma'$
to electrons in thermal plasma. }
\end{figure}

During this epoch, dark photons are dominantly produced through thermal
processes involving $\gamma$ and $e^{\pm}$, including (i) inverse
decay ($e^{+}e^{-}\to\gamma'$), (ii) electron-positron annihilation
($e^{+}e^{-}\to\gamma'\gamma$), and (iii) semi-Compton scattering
($\gamma e^{\pm}\to\gamma'e^{\pm}$), as shown in Fig.~\ref{fig:feyn}.
The inverse decay process is kinematically allowed only when $m_{\gamma'}>2m_{e}$.
The annihilation process is Boltzmann suppressed when $T$ drops below
$m_{e}$. The semi-Compton process is less suppressed at low temperatures
compared to annihilation, and becomes the most important process at
$T\lesssim20\ \text{keV}$, below which positrons almost disappear
while a certain amount of electrons remain due to  matter-antimatter
asymmetry. 

Below, we calculate the collision terms of these processes. Note that
the calculation in this section does not include medium effects and
the resulting production rates may differ significantly from the actual
rates in medium. Only when the plasma frequency is well below $m_{\gamma'}$,
the production rates obtained in this section are approximately equal
to the in-medium ones---see Eq.~\eqref{eq:-16} in the next section.
For simplicity, we assume Boltzmann statistics, which is a common
approximation used in collision term calculations.\footnote{See the textbook by Kolb \& Turner \cite{Kolb} for discussions on
the validity of this approximation. Generally speaking, the difference
caused by using the Boltzmann approximation is small when quantum
statistical effects (Bose condensation or Pauli blocking) are insignificant.
In our work, this is always the case. The collision terms computed
with the Boltzmann approximation typically deviate from the exact
results by ${\cal O}(10\%)$---see, e.g., Eqs.~(A.5)-(A.6) in \cite{EscuderoAbenza:2020cmq}
or Tab.~III in \cite{Luo:2020sho}.}   

\subsection{Inverse decay}

For the inverse decay process, $e^{+}e^{-}\to\gamma'$, the gain rate
of $\gamma'$ reads
\begin{align}
\Gamma_{e^{+}e^{-}\to\gamma'}^{{\rm gain}} & =\frac{1}{2\omega}\int g_{1}d\Pi_{1}g_{2}d\Pi_{2}f_{1}f_{2}(2\pi\delta)^{4}|{\cal M}|^{2},\label{eq:}
\end{align}
where $\omega$ denotes the energy of $\gamma'$, $d\Pi_{i}=\frac{d^{3}p_{i}}{(2\pi)^{3}2E_{i}}$
with $p_{i}$ and $E_{i}$ the momentum and energy  of the $i$-th
particle in the considered process, $(2\pi\delta)^{4}$ is the common
delta function responsible for momentum conservation, and $|{\cal M}|^{2}$
is the squared matrix element (with all spins and polarizations averaged
out). The factor $g_{i}$ accounts for the multiplicity of particle
$i$, and $f_{i}$ denotes its phase space distribution. 

The squared matrix element of $e^{+}e^{-}\to\gamma'$ is given by
\begin{equation}
|{\cal M}|^{2}=\frac{1}{g_{\gamma'}g_{e^{+}}g_{e^{-}}}4e^{2}\varepsilon^{2}\left(m_{\gamma'}^{2}+2m_{e}^{2}\right),\label{eq:M2-decay}
\end{equation}
where $g_{e^{+}}=g_{e^{-}}=2$ account for the two degrees of freedom
of spin. They will be canceled out by $g_{1}$ and $g_{2}$ in Eq.~\eqref{eq:}.
Similarly, the multiplicity factor $g_{\gamma'}$ will eventually
be canceled out by the $g_{\gamma'}$ factor in Eq.~\eqref{eq:Gamma-to-C}.

Since $|{\cal M}|^{2}$ in Eq.~\eqref{eq:M2-decay} is a constant,
we may pull it outside the integral and perform the phase space integration
straightforwardly. The technical details of computing such two-body
phase space integrals are presented in Appendix~\ref{sec:two-body-integral}.
The result reads
\begin{align}
\Gamma_{e^{+}e^{-}\to\gamma'}^{{\rm gain}} & =\alpha\varepsilon^{2}\frac{m_{\gamma'}^{2}+2m_{e}^{2}}{3\omega}\sqrt{1-\frac{4m_{e}^{2}}{m_{\gamma'}^{2}}}e^{-\omega/T}.\label{eq:-1}
\end{align}
Using Eq.~\eqref{eq:Gamma-to-C}, we obtain
\begin{align}
C_{e^{+}e^{-}\to\gamma'}^{{\rm gain}} & =\alpha\varepsilon^{2}\frac{m_{\gamma'}^{2}+2m_{e}^{2}}{2\pi^{2}}m_{\gamma'}T\sqrt{1-\frac{4m_{e}^{2}}{m_{\gamma'}^{2}}}K_{1}\left(\frac{m_{\gamma'}}{T}\right),\label{eq:-2}
\end{align}
where $K_{1}$ is the modified Bessel function.  The above results
are included in Tab.~\ref{tab:Collision-terms}. Eq.~\eqref{eq:-2}
has been previously computed in Refs.~\cite{Redondo:2008ec,Fradette:2014sza,Gan:2023wnp},
and our result agrees with those in the literature. The squared matrix
element has been explicitly presented in \cite{Fradette:2014sza},
in the form that all initial and final state spin degrees of freedom
are summed up. Hence it corresponds to our Eq.~\eqref{eq:M2-decay}
without the prefactor $1/(g_{\gamma'}g_{e^{+}}g_{e^{-}})$.

\begin{table}
\centering

\renewcommand{\arraystretch}{2.1} 

\begin{tabular}{ccc}
\hline 
 & $\Gamma_{\gamma'}^{{\rm gain}}/\varepsilon^{2}$ & $C_{\gamma'}^{{\rm gain}}/\varepsilon^{2}$\tabularnewline
\hline 
$e^{+}e^{-}\to\gamma'$ & $\alpha\frac{m_{\gamma'}^{2}+2m_{e}^{2}}{3\omega}\sqrt{1-\frac{4m_{e}^{2}}{m_{\gamma'}^{2}}}e^{-\omega/T}$ & $\alpha\frac{m_{\gamma'}^{2}+2m_{e}^{2}}{2\pi^{2}}m_{\gamma'}T\sqrt{1-\frac{4m_{e}^{2}}{m_{\gamma'}^{2}}}K_{1}\left(\frac{m_{\gamma'}}{T}\right)$\tabularnewline
\makecell{$e^{+}e^{-}\to\gamma'\gamma$\\[-0.2em] $(T\gtrsim m_{e})$} & $\frac{\alpha^{2}g_{\Gamma}T^{2}}{6\pi\omega}\left[\log\left(\frac{4T\omega}{m_{e}^{2}}\right)-\gamma_{E}-1\right]e^{-\frac{\omega}{T}}$ & $\frac{\alpha^{2}g_{C}T^{4}}{12\pi^{3}}\left[\log\left(\frac{4T^{2}}{m_{e}^{2}}\right)-2\gamma_{E}\right]$\tabularnewline
\makecell{$e^{+}e^{-}\to\gamma'\gamma$\\[-0.2em] $(T\lesssim m_{e})$} & $\frac{\alpha^{2}g_{\Gamma}T^{5/2}}{12\sqrt{\pi\omega}m_{e}}e^{-\frac{m_{e}^{2}+\omega^{2}}{T\omega}}$ & $\alpha^{2}g_{C}\frac{4m_{e}^{2}T^{3}+6m_{e}T^{4}+3T^{5}}{96\pi^{2}m_{e}}e^{-\frac{2m_{e}}{T}}$\tabularnewline
\makecell{$\gamma e^{\pm}\to\gamma'e^{\pm}$\\[-0.2em] $(T\gtrsim m_{e})$} & $\frac{\alpha^{2}g_{\Gamma}T^{2}}{12\pi\omega}\left[\log\left(\frac{4T\omega}{m_{e}^{2}}\right)-\gamma_{E}+\frac{1}{2}\right]e^{-\frac{\omega}{T}}$ & $\frac{\alpha^{2}g_{C}T^{4}}{12\pi^{3}}\left[\log\left(\frac{2T}{m_{e}}\right)-\gamma_{E}+\frac{3}{4}\right]$\tabularnewline
\makecell{$\gamma e^{\pm}\to\gamma'e^{\pm}$\\[-0.2em] $(T\lesssim m_{e})$} & $\frac{\sqrt{2}\alpha^{2}g_{\Gamma}T^{2}}{9\sqrt{\pi Tm_{e}}}\exp\left(\frac{\mu_{e^{\pm}}-m_{e}-\omega}{T}\right)$ & $\frac{\sqrt{2}\alpha^{2}g_{C}T^{9/2}}{9\pi^{2}\sqrt{\pi m_{e}}}\exp\left(\frac{\mu_{e^{\pm}}-m_{e}}{T}\right)$\tabularnewline
\hline 
\end{tabular}\caption{\label{tab:Collision-terms} Collision terms responsible for the production
of $\gamma'$. For $e^{+}e^{-}\to\gamma'$, the presented expressions
are exact. For $e^{+}e^{-}\to\gamma'\gamma$ and $\gamma e^{-}\to\gamma'e^{-}$,
the expressions are derived with assumptions that high-$T$ or low-$T$
expansions are valid and $m_{\gamma'}$ is negligibly small.  A few
constants are defined as follows: $g_{\Gamma}\equiv g_{e^{-}}g_{e^{+}}g_{\gamma}=8$,
$g_{C}\equiv g_{e^{-}}g_{e^{+}}g_{\gamma}g_{\gamma'}=24$, and $\gamma_{E}$
is Euler's constant. The accuracy of these approximate expressions
is demonstrated in Fig.~\ref{fig:MC}.}
\end{table}

\subsection{Annihilation and semi-Compton}

For two-to-two processes like $e^{+}e^{-}\to\gamma\gamma'$ and $\gamma e^{\pm}\to e^{\pm}\gamma'$,
it is generally difficult to work out the phase space integration
analytically but, with certain approximations, we are able to derive
analytical expressions---see Appendix~\ref{sec:Collision-terms}
for the derivation and Tab.~\ref{tab:Collision-terms} for a summary
of the results. On the other hand, using the Monte Carlo method described
 in Appendix~B of \cite{Luo:2020fdt}, we also perform the integration
numerically. 

\begin{figure}
\centering

\includegraphics[width=0.48\textwidth]{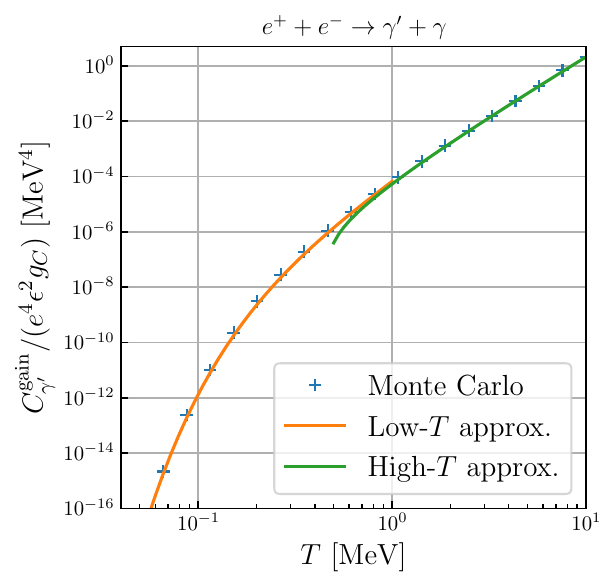}\includegraphics[width=0.48\textwidth]{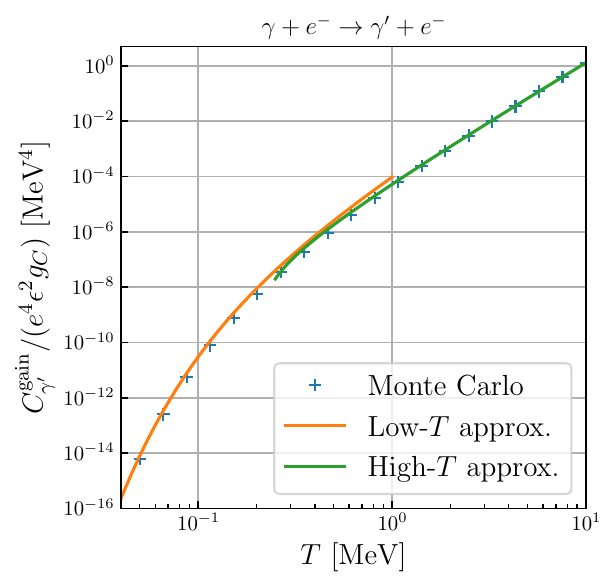}

\caption{\label{fig:MC} Comparison of the analytical expressions for two-to-two
collision terms  with Monte-Carlo results. }
\end{figure}

In Fig.~\ref{fig:MC}, we compare the analytical expressions with
the Monte Carlo results. Here, the green and orange lines represent
analytical expressions derived using the high-$T$ and low-$T$ approximations,
respectively.  These approximations are expected to be  valid for
$T\gg m_{e}$ or $T\ll m_{e}$, but in practice one can see that the
resulting expressions remain accurate even when $T$ is comparable
to $m_{e}$. The comparison in Fig.~\ref{fig:MC} suggests that the
high-$T$ and low-$T$ expressions can be matched to produce an expression
that remains accurate across the full range, with the matching point
set at $T=2m_{e}$ for $e^{+}e^{-}\to\gamma\gamma'$ , and $T=m_{e}$
for $\gamma e^{\pm}\to e^{\pm}\gamma'$. 

It should be noted that when $T$ is well below $m_{e}$, the electron-positron
asymmetry becomes important, as it stops the electron number density
from being further Boltzmann suppressed. At $T\lesssim20\ \text{keV}$,
most positrons have annihilated with electrons while a residual population
of electrons survives due to the asymmetry. Consequently, once the
asymmetry becomes important, processes involving positrons effectively
cease, and the production of $\gamma'$ proceeds only through $\gamma e^{-}\to e^{-}\gamma'$.
To include the asymmetry effect, we introduce chemical potentials
$\mu_{e^{\pm}}$ for $e^{\pm}$ in the last row of Tab.~\ref{tab:Collision-terms},
i.e., the expressions with the low-$T$ approximation of the semi-Compton
process. For other collision terms, the asymmetry  effect is negligible. 

We note here that the low-$T$ collision term for the semi-Compton
process can be derived simply using the Thomson scattering cross section
multiplied by the electron number density, assuming that the electron
is at rest. This has been previously calculated in Refs.~\cite{Pospelov:2008jk,Redondo:2008ec}.
The high-$T$ collision term for this process can be found in Ref.~\cite{Redondo:2008ec},
in which its damping factor corresponds to $\Gamma_{\gamma'}^{{\rm gain}}(1+f_{\gamma})/f_{\gamma}$
in our work. Both the low- and high-$T$ results in Ref.~\cite{Redondo:2008ec}
approximately agree with ours. The collision term for the annihilation
process has been previously calculated in \cite{Redondo:2008ec},
but only numerically. Hence, Tab.~\ref{tab:Collision-terms} represents
the most complete analytical calculation of relevant collision terms
for dark photon production in the early-universe plasma.

\subsection{High-mass-regime yield estimate \label{subsec:High-mass-regime-yield}}

When the dark photon is relatively heavy (compared with $m_{e}$),
the dominant production channel is $e^{+}e^{-}\to\gamma'$, since
$e^{+}e^{-}\to\gamma\gamma'$ and $\gamma e^{\pm}\to e^{\pm}\gamma'$
are suppressed by an extra power of $\alpha$. Given the simplicity
of the collision term $C_{e^{+}e^{-}\to\gamma'}^{{\rm gain}}$, it
is straightforward to plug it into Eq.~\eqref{eq:n-int} and obtain
\begin{equation}
\frac{n_{\gamma'}}{n_{\gamma}}\approx\frac{\pi\alpha\varepsilon^{2}m_{{\rm pl}}}{4\zeta(3)g_{H}m_{\gamma'}}\frac{g_{\gamma'}}{g_{\gamma}}\left(1+2\frac{m_{e}^{2}}{m_{\gamma'}^{2}}\right)\sqrt{1-\frac{4m_{e}^{2}}{m_{\gamma'}^{2}}}\thinspace,\label{eq:-41}
\end{equation}
where $\zeta$ is the Riemann Zeta function. Equation~\eqref{eq:-41}
offers a very useful estimate of the yield of heavy dark photons in
the freeze-in regime. It should be interpreted as the ratio of $n_{\gamma'}$
to $n_{\gamma}$ when the production has effectively ceased while
the decay has not started yet. In the freeze-in regime, such a period
always exists. The value of $g_{H}$ should be evaluated at $T\sim m_{\gamma'}$,
assuming that $g_{\star}$ does not vary rapidly when $T$ approaches
$m_{\gamma'}$. 

The validity of Eq.~\eqref{eq:-41} only requires that $m_{\gamma'}$
is above $2m_{e}$ and $\varepsilon$ is sufficiently small.  It
remains valid even after including the medium effect which is responsible
for  the resonant production---see discussions in the next section
and Fig.~\ref{fig:sol}. 

Neglecting the factor $\left(1+2m_{e}^{2}/m_{\gamma'}^{2}\right)\sqrt{1-4m_{e}^{2}/m_{\gamma'}^{2}}$,
which only deviates from $1$ by less than $10\%$ for $m_{\gamma'}>3m_{e}$,
Eq.~\eqref{eq:-41} implies
\begin{equation}
\frac{n_{\gamma'}}{n_{\gamma}}\approx0.011\times\frac{g_{\gamma'}}{g_{\gamma}}\cdot\left(\frac{\varepsilon}{10^{-10}}\right)^{2}\cdot\frac{10\ \text{MeV}}{m_{\gamma'}}\cdot\left(\frac{10.75}{g_{\star}}\right)^{1/2}.\label{eq:-42}
\end{equation}
The freeze-in estimate would lose its validity when $n_{\gamma'}$
approaches $n_{\gamma}$. Hence for the above expression to be valid,
$\varepsilon$ should satisfy $\varepsilon\lesssim10^{-9}\sqrt{m_{\gamma'}/10\ \text{MeV}}$.

\section{The medium effect\label{sec:medium-effect}}

In the thermal environment of the early universe, the photon receives
a thermal correction to its self-energy from coherent scattering with
the medium particles. This thermal correction, also known as the thermal
mass of the photon, modifies the derivation of Eq.~\eqref{eq:J-A-A},
which  is derived from the diagonalization of the kinetic and mass
terms of gauge bosons.  Note that $\gamma$ and $\gamma'$ are defined
as mass eigenstates with canonically-normalized kinetic terms.  Therefore,
in the medium when the photon attains the thermal mass, these mass
eigenstates, together with the derivation of Eq.~\eqref{eq:J-A-A},
 are altered by the medium effect.  

The medium effect is small when $m_{\gamma'}$ is large compared to
the thermal effective mass of $\gamma$. However, when $m_{\gamma'}$
is small or the medium becomes sufficiently dense and hot, the influence
of the medium effect rises.  In particular, if $m_{\gamma'}\to0$,
the dark photon in the medium would behave as if it is fully decoupled
from the SM sector. This is the reason why most cosmological and astrophysical
bounds on the dark photon vanish in the limit of $m_{\gamma'}\to0$~\cite{Caputo:2021eaa,AxionLimits}.

\subsection{Basic formalism }

The medium effect can be included by employing the effective in-medium
mixing $\varepsilon_{\text{m}}$, which is related to the in-vacuum
mixing $\varepsilon$ as~\cite{Redondo:2008aa,An:2013yfc,Redondo:2013lna,Li:2023vpv}
\begin{equation}
\varepsilon_{\text{m}}=\frac{m_{\gamma'}^{2}}{m_{\gamma'}^{2}-\Pi_{\gamma\gamma}}\varepsilon\thinspace,\label{eq:epsilon-m}
\end{equation}
where $\Pi_{\gamma\gamma}$ denotes the photon self-energy in the
thermal environment. Eq.~\eqref{eq:epsilon-m} can be derived in various
approaches---see, e.g., Eq.~(15) of Ref.~\cite{Redondo:2008aa}
for a derivation from the classical equation of motion, or Fig.~1
and Eq.~(3.7) of Ref.~\cite{Li:2023vpv} for a diagrammatic derivation. 

Including the medium effect, the production rate of $\gamma'$ in
the thermal bath is given by
\begin{equation}
\Gamma_{\gamma'}^{{\rm gain}}=\frac{m_{\gamma'}^{4}}{\left(m_{\gamma'}^{2}-\text{Re}\Pi_{\gamma\gamma}\right)^{2}+\left(\text{Im}\Pi_{\gamma\gamma}\right)^{2}}\Gamma_{\gamma',\text{vac}}^{{\rm gain}}\thinspace,\label{eq:-16}
\end{equation}
where $\text{Re}\Pi_{\gamma\gamma}$ and $\text{Im}\Pi_{\gamma\gamma}$
are the real and imaginary parts of $\Pi_{\gamma\gamma}$, and $\Gamma_{\gamma',\text{vac}}^{{\rm gain}}$
is the production rate of $\gamma'$ without including the medium
effect. Diagrammatically, $\Gamma_{\gamma'}^{{\rm gain}}$ and $\Gamma_{\gamma',\text{vac}}^{{\rm gain}}$
correspond to the Feynman diagrams in Fig.~\ref{fig:feyn} with the
hatched blobs equal to $\varepsilon_{\text{m}}e$ and $\varepsilon e$,
respectively.    

The real part $\text{Re}\Pi_{\gamma\gamma}$ in certain limits has
simple analytical results but in general involves an integral that
requires numerical integration~\cite{Braaten:1993jw}. It is noteworthy
that the integral becomes sensitive to the electron asymmetry  at
low temperatures below around $20$ keV. In Appendix~\ref{sec:Pi-gg},
we briefly review the calculation of $\text{Re}\Pi_{\gamma\gamma}$
based on Ref.~\cite{Braaten:1993jw}. 

The imaginary part $\text{Im}\Pi_{\gamma\gamma}$ is related to the
photon interaction rate $\Gamma_{\gamma}$ by~\cite{Weldon:1983jn}
\begin{equation}
\text{Im}\Pi_{\gamma\gamma}=-\omega\Gamma_{\gamma}\thinspace,\label{eq:-14}
\end{equation}
with 
\begin{equation}
\Gamma_{\gamma}\equiv\Gamma_{\gamma}^{{\rm loss}}-\Gamma_{\gamma}^{{\rm gain}}\thinspace.\label{eq:-12}
\end{equation}
Here $\Gamma_{\gamma}^{{\rm gain}}$ and $\Gamma_{\gamma}^{{\rm loss}}$
are defined as the gain and loss rates of $\gamma$ in a way similar
to $\Gamma_{\gamma'}^{{\rm gain}}$ and $\Gamma_{\gamma'}^{{\rm loss}}$
defined in Eq.~\eqref{eq:dfdt}. Similar to Eq.~\eqref{eq:dfdt},
one could also write down the Boltzmann equation for $f_{\gamma}$.
However,  since $\gamma$ is in thermal equilibrium, the right-hand
side of this equation should vanish, i.e.,
\begin{equation}
\Gamma_{\gamma}^{{\rm gain}}\left(1+f_{\gamma}\right)-\Gamma_{\gamma}^{{\rm loss}}f_{\gamma}=0\thinspace,\label{eq:-13}
\end{equation}
where $f_{\gamma}=1/\left(e^{\omega/T}-1\right)$ is Bose-Einstein
distribution of photons in thermal equilibrium. Viewing Eqs.~\eqref{eq:-12}
and \eqref{eq:-13} as linear equations in $\Gamma_{\gamma}^{{\rm gain}}$
and $\Gamma_{\gamma}^{{\rm loss}}$, one can solve them to obtain
\begin{equation}
\Gamma_{\gamma}^{{\rm gain}}=f_{\gamma}\Gamma_{\gamma}\thinspace,\ \ \Gamma_{\gamma}^{{\rm loss}}=(1+f_{\gamma})\Gamma_{\gamma}\thinspace.\label{eq:-15}
\end{equation}
Eq.~\eqref{eq:-15} allows us to determine $\Gamma_{\gamma}$ from
$\Gamma_{\gamma}^{{\rm gain}}$. The latter is determined by Feynman
diagrams similar to  those in Fig.~\ref{fig:feyn}. So the results
can be obtained by adapting those in Tab.~\ref{tab:Collision-terms}
with $\varepsilon\to1$ and $m_{\gamma'}^{2}\to\omega^{2}-k^{2}$.
Although this includes the contribution of $e^{+}e^{-}\to\gamma$,
it does not imply that $\gamma$ can be physically produced via this
process, as the on-shell production of such photons would require
$\omega^{2}-k^{2}>4m_{e}^{2}$, which is kinematically forbidden for
photons\footnote{Even at high temperatures when the thermal mass of $\gamma$ exceeds
$2m_{e}$, this production channel remains kinematically forbidden
because the dispersion relation of the electron is also modified~\cite{Braaten:1993jw}.
More specifically, the electron thermal mass squared in the high-$T$
limit, according to Eq.~(6.131) in Ref.~\cite{Bellac}, is $e^{2}T^{2}/8\approx0.01T^{2}$,
which is still higher than the high-$T$ limit of the plasma frequency,
$\omega_{P}^{2}=e^{2}T^{2}/9$, implying that $\gamma\to e^{+}e^{-}$
remains kinematically forbidden even at high temperatures. One might
be concerned about whether the thermal correction to the electron
dispersion relation could affect our analysis. In fact, this influence
is very limited because the high-temperature production of dark photons
is important only for heavy dark photons, which are predominantly
produced via $e^{+}e^{-}\to\gamma'$ at $T\sim m_{\gamma'}/3$. At
this temperature, the ratio of mass squared between electrons and
dark photons is $0.01T^{2}/(3T)^{2}\approx10^{-3}$, which is negligibly
small. Hence, the influence of the electron thermal mass can be safely
neglected in our analysis.}.  The on-shell production of $\gamma'$ via $e^{+}e^{-}$ coalescence,
however, is possible when $m_{\gamma'}^{2}>4m_{e}^{2}$ and becomes
the dominant channel in the high-mass regime. When the production
of $\gamma'$ is significantly affected by the contribution of $e^{+}e^{-}\to\gamma$
to $\Gamma_{\gamma}^{{\rm gain}}$ (This occurs at $T\to\frac{3}{2\sqrt{\pi\alpha}}m_{\gamma'}\approx9.9m_{\gamma'}$
 for $\gamma'$ heavier than $2m_{e}$), it can be physically interpreted
as that $e^{+}e^{-}$ coalescence first produces an off-shell $\gamma$
which is then converted, via kinetic mixing, to an on-shell $\gamma'$
with the same $\omega$ and $k$. 

Using Eq.~\eqref{eq:-16} with $\text{Re}\Pi_{\gamma\gamma}$ and
$\text{Im}\Pi_{\gamma\gamma}$ evaluated as described above, it is
straightforward to calculate $\Gamma_{\gamma'}^{{\rm gain}}$,  at
least numerically. Nevertheless, in the next two subsections, we would
like to discuss some analytical features which, while not essential
for the numerical evaluation, provide important  insight into the
thermal production of dark photons.  

\subsection{Resonant production\label{subsec:Resonant}}

When $\text{Re}\Pi_{\gamma\gamma}$ in Eq.~\eqref{eq:-16} approaches
$m_{\gamma'}^{2}$, the denominator reduces to $(\text{Im}\Pi_{\gamma\gamma})^{2}$,
indicating that the production of $\gamma'$ is greatly enhanced,
i.e., it enters the resonant production regime. When ${\rm Im}\Pi_{\gamma\gamma}$
is small, the resonant production can be approximated by the Dirac
delta function:
\begin{equation}
\Gamma_{\gamma'}^{{\rm gain}}\approx f_{\gamma}\Gamma_{\gamma}\frac{\varepsilon^{2}m_{\gamma'}^{4}}{\left(m_{\gamma'}^{2}-\text{Re}\Pi_{\gamma\gamma}\right)^{2}+\omega^{2}\Gamma_{\gamma}^{2}}\xrightarrow{\text{res.}}f_{\gamma}\varepsilon^{2}m_{\gamma'}^{4}\frac{\pi}{\omega}\delta\left(m_{\gamma'}^{2}-\text{Re}\Pi_{\gamma\gamma}\right),\label{eq:-17}
\end{equation}
where we have used the identity $\lim_{y\to0}\ \frac{y}{x^{2}+y^{2}}=\pi\delta(x)$.
The abbreviation ``res.'' above the arrows indicates that this is
an approximation made near the resonance. Note that the result after
taking the $\delta$ function limit becomes independent of $\Gamma_{\gamma}$.
As a consequence, if we integrate  the  production rate around the
resonance, the resulting yield of dark photons is insensitive to $\Gamma_{\gamma}$. 

Substituting Eq.~\eqref{eq:-17} into Eq.~\eqref{eq:f-int}, we obtain
\begin{align}
f_{\gamma'}(a,\ p) & \approx\frac{\pi\varepsilon^{2}m_{\gamma'}^{4}}{apH_{\star}}\left|\frac{d\text{Re}\Pi_{\gamma\gamma}}{da}\right|_{\star}^{-1}\exp\left(-\frac{ap}{a_{\star}T_{\star}}\right),\label{eq:-18}
\end{align}
where we have used the Boltzmann approximation and the relativistic
approximation. Here the subscript ``$\star$'' indicates that the
quantity should be evaluated at the resonance.  Integrating over
the phase space,  we obtain
\begin{equation}
n_{\gamma'}\approx\frac{\varepsilon^{2}g_{\gamma'}m_{\gamma'}^{4}a_{\star}^{2}T_{\star}^{2}}{2\pi a^{3}H_{\star}}\left|\frac{d\text{Re}\Pi_{\gamma\gamma}}{da}\right|_{\star}^{-1},\label{eq:-19}
\end{equation}
 where we have assumed transverse polarizations such that the mild
energy dependence in $\text{Re}\Pi_{\gamma\gamma}$ can be neglected. 

Using a semi-analytic expression for $\text{Re}\Pi_{\gamma\gamma}$
in Appendix~\ref{sec:Pi-gg}, we can further write Eq.~\eqref{eq:-19}
as 
\begin{equation}
n_{\gamma'}\approx\left(\frac{a_{\star}}{a}\right)^{3}\cdot\frac{9\varepsilon^{2}g_{\gamma'}m_{\gamma'}^{4}}{16\pi^{2}\alpha H_{\star}}F_{1}\left(\frac{T_{\star}}{\text{MeV}}\right),\label{eq:-20}
\end{equation}
where
\begin{equation}
F_{1}(x)\approx\frac{1}{\frac{3}{2}xc_{5}^{2}+\left(\frac{2}{15}x^{-\frac{6}{5}}+1\right)e^{-\frac{2}{9x^{6/5}}}}\thinspace,\label{eq:-21}
\end{equation}
and $c_{5}\approx5.0\times10^{-5}$ is a constant determined by the
electron asymmetry. Note that $F_{1}(x)\approx1$ for large $x$.
We emphasize here that Eq.~\eqref{eq:-20} is only applicable to transverse
polarizations. The longitudinal contribution requires numerical integration
of $p$ and is known to be suppressed~\cite{Fradette:2014sza}, so
we only include it in the numerical calculation. 

Comparing $n_{\gamma'}$ to the photon number density $n_{\gamma}$,
we find 
\begin{equation}
\frac{n_{\gamma'}}{n_{\gamma}}\approx\frac{9\varepsilon^{2}g_{\gamma'}m_{\gamma'}^{4}m_{{\rm pl}}}{16\zeta(3)\alpha g_{\gamma}g_{H}T_{\star}^{5}}F_{1}\left(\frac{T_{\star}}{\text{MeV}}\right).\label{eq:-22}
\end{equation}
This result should be understood as the value of $n_{\gamma'}/n_{\gamma}$
shortly after the resonant production. Subsequent evolution could
modify it via, e.g., dark photon decay or entropy dilution. 

If the resonance occurs when electrons are still relativistic ($T_{\star}\gtrsim m_{e}$),
Eq.~\eqref{eq:-22} reduces to
\begin{equation}
\frac{n_{\gamma'}}{n_{\gamma}}\approx0.015\times\left(\frac{\varepsilon}{10^{-10}}\right)^{2}\cdot\frac{1\ \text{MeV}}{m_{\gamma'}}\cdot\left(\frac{10.75}{g_{\star}}\right)^{1/2},\label{eq:-23}
\end{equation}
which is a simple and convenient formula for estimating the freeze-in
yield of  resonant production. In practice, we find that Eq.~\eqref{eq:-23}
is valid for $m_{\gamma'}\gtrsim0.05\text{MeV}$.  For lower masses,
we recommend using Eq.~\eqref{eq:-22} with Eq.~\eqref{eq:-21} to
estimate the ratio. 

After the resonance, the production of $\gamma'$ proceeds at a rate
without significant enhancement from  the medium effect (In particular,
when $\text{Re}\Pi_{\gamma\gamma}\ll m_{\gamma'}^{2}$, the rate is
well approximated by $\Gamma_{\gamma',\text{vac}}^{{\rm gain}}$).
However, this does not necessarily mean that the resonant production
makes the dominant contribution throughout the entire evolution. In
the high-mass regime with $m_{\gamma'}>2m_{e}$, the resonance occurs
at $T\approx9.9m_{\gamma'}$ but at $T\sim m_{\gamma'}$ the inverse
decay process also contributes significantly, as we have estimated
in Sec.~\ref{subsec:High-mass-regime-yield}.   It is therefore
interesting to compare Eq.~\eqref{eq:-23} or \eqref{eq:-22} with
Eq.~\eqref{eq:-42} or \eqref{eq:-41}. Note that  Eq.~\eqref{eq:-22}
is proportional to $\alpha^{3/2}$ because $T_{\star}\propto\alpha^{-1/2}$
and $F_{1}\approx1$, while Eq.~\eqref{eq:-41} is proportional to
$\alpha$. Hence we expect that the resonant production is subdominant
in the high-mass regime.  Indeed, if we compute the ratio of Eq.~\eqref{eq:-22}
to \eqref{eq:-41}, we find
\begin{equation}
\frac{\left(n_{\gamma'}a^{3}\right)_{{\rm res}.}}{\left(n_{\gamma'}a^{3}\right)_{{\rm ID}}}\approx\frac{4\pi e}{27}+{\cal O}\left(\frac{m_{e}^{4}}{m_{\gamma'}^{4}}\right)\approx0.14\thinspace.\label{eq:-59}
\end{equation}
Here the subscript ``res.'' and ``ID'' indicate the contributions
from the resonant production  and from the post-resonance inverse
decay process.  In Sec.~\ref{sec:Cosmological-evolution}, we will
show that Eq.~\eqref{eq:-59} agrees very well with numerical solutions---see
Fig.~\ref{fig:sol}. We note here that this subdominant contribution
has already been calculated in Ref.~\cite{Fradette:2014sza} in which
the integration of $\omega$ was performed numerically due to the
complexity of the full expression of $\text{Re}\Pi_{\gamma\gamma}$.
We have checked that if the approximate expression of $\text{Re}\Pi_{\gamma\gamma}$
used here is substituted into the formalism in Ref.~\cite{Fradette:2014sza},
the same result can be obtained.

\subsection{Photon-dark photon oscillation }

Although the photon and dark photon are in a diagonal basis after
their kinetic terms are canonically diagonalized in vacuum {[}i.e.,
the step from Eq.~\eqref{eq:FF} to Eq.~\eqref{eq:J-A-A}{]}, the
medium effect reintroduces an effective mixing between them. Consequently,
when photons propagate from vacuum to medium, or through medium with
varying density, photon-dark photon oscillation can occur. This is
analogous to neutrino oscillations affected by the matter effect~\cite{Wolfenstein:1977ue,Mikheev:1986gs,Mikheev:1986wj},
or to photon-axion conversion (or oscillation) in magnetic fields---see,
e.g., \cite{Wu:2024fsf}. 

The oscillation length of $\gamma\leftrightarrow\gamma'$ is of the
order $\omega/\Delta m^{2}$ where $\Delta m^{2}$ denotes the mass
squared difference between the two particles in medium\footnote{Note that in medium, this difference can significantly deviate from
$m_{\gamma'}^{2}$. In particular, if it is near the resonance, it
becomes suppressed by $\varepsilon$---see discussions below Eq.~\eqref{eq:-64}. }. This length is to be compared with the mean free path of $\gamma$,
given by $1/\Gamma_{\gamma}^{{\rm loss}}\sim1/\Gamma_{\gamma}$. In
the weak-damping regime ($\omega/\Delta m^{2}\ll1/\Gamma_{\gamma}$),
 the validity of oscillation is obvious, as a quantum superposition
state of $\gamma$ and $\gamma'$ has sufficient space to oscillate
before it is absorbed by the medium. In the strong-damping regime
($\omega/\Delta m^{2}\gg1/\Gamma_{\gamma}$) which is also referred
to as the quantum Zeno regime in Ref.~\cite{Redondo:2013lna}, the
oscillation formalism can still be used to describe the production
of $\gamma'$ because frequent scattering in this case plays the role
of frequent measurements in the quantum Zeno effect. In this subsection
we will show that indeed the resonant production given by Eq.~\eqref{eq:-17}
can be reproduced in the oscillation formalism assuming constant medium.
 A full treatment of photon-dark photon oscillation in the expanding
universe, where the medium varies with time (one can also regard it
as variation along the propagation path), is rather involved and  not
necessary here, since within its validity range it is equivalent to
the calculation presented above. Below we demonstrate some illuminating
features in the oscillation framework that may deepen our understanding
of the dark-photon production in thermal medium. 

\begin{figure}
\centering

\includegraphics[width=0.5\textwidth]{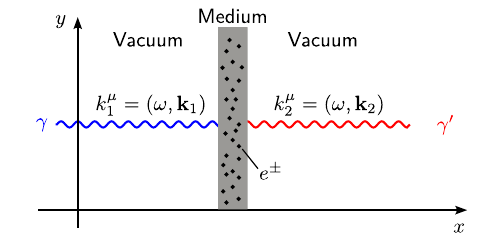}

\caption{\label{fig:coh} Schematic illustration of the thin-slab approximation
used to derive $\gamma$-$\gamma'$ oscillation.}
\end{figure}

Let us consider a simple scenario in which a photon propagates from
vacuum, through a thin slab of medium, and then back into vacuum,
as illustrated in Fig.~\ref{fig:coh}.  In vacuum, the photon can
never be converted to a dark photon since the kinetic mixing has been
eliminated when canonically diagonalizing the kinetic terms. When
it propagates through the slab of medium, there is a small probability
that such conversion can happen, since the medium generates an effective
operator,\footnote{See e.g.~Appendix B in \cite{Li:2023vpv} for an explicit calculation
of the coherent scattering with medium particles that generates the
operator.} ${\cal L}\supset\Pi_{\gamma\gamma'}A^{\mu}A'_{\mu}$ where $\Pi_{\gamma\gamma'}=\varepsilon\Pi_{\gamma\gamma}$.
Then the following matrix element gives rise to the conversion amplitude:\begin{equation}
\langle
 \wick{
  \c1{\mathbf{k}_{2}}| \Pi_{\gamma \gamma'} \c2 A^{\mu} \c1 A'_{\mu} |\c2{\mathbf{k}_{1}}
  }
\rangle \to {\cal A}_{\gamma\to\gamma'}\,.
\label{eq:wick}
\end{equation}Here $|\mathbf{k}_{1}\rangle$ and $|\mathbf{k}_{2}\rangle$ denote
the initial and final states of $\gamma$ and $\gamma'$, respectively.
Note that, as indicated in Fig.~\ref{fig:coh}, they share the same
energy $\omega$ but their momenta are different, which implies that
there is a small momentum transfer to/from the medium, a well-known
aspect in photon refraction. 

After performing the Wick contraction, we obtain\footnote{See Eq.~(A.21) in \cite{Wu:2024fsf} for a derivation.}
\begin{equation}
{\cal A}_{\gamma\to\gamma'}=\frac{1}{2\sqrt{|\mathbf{k}_{1}||\mathbf{k}_{2}|}}\int dx\thinspace\Pi_{\gamma\gamma'}(x)\thinspace e^{i|\mathbf{k}_{1}-\mathbf{k}_{2}|x}\thinspace.\label{eq:-24}
\end{equation}
The integral in Eq.~\eqref{eq:-24} is essentially a Fourier transform
of $\Pi_{\gamma\gamma'}(x)$ which we assume is a constant in the
slab, i.e.,
\begin{equation}
\Pi_{\gamma\gamma'}(x)=\begin{cases}
\Pi_{\gamma\gamma'} & x\in[0,\Delta L]\\
0 & {\rm otherwise}
\end{cases}\thinspace.\label{eq:-27}
\end{equation}
Here $\Delta L$ is the thickness of the slab. 

  Substituting Eq.~\eqref{eq:-27} into Eq.~\eqref{eq:-24}, we
get
\begin{equation}
{\cal A}_{\gamma\to\gamma'}=\frac{\Pi_{\gamma\gamma'}}{2\sqrt{|\mathbf{k}_{1}||\mathbf{k}_{2}|}}\left(\frac{e^{i|\mathbf{k}_{1}-\mathbf{k}_{2}|\Delta L}-1}{|\mathbf{k}_{1}-\mathbf{k}_{2}|}\right),\label{eq:-28}
\end{equation}
which in the thin-slab limit gives
\begin{equation}
\lim_{\Delta L\to0}\frac{{\cal A}_{\gamma\to\gamma'}}{\Delta L}=i\frac{\Pi_{\gamma\gamma'}}{2\sqrt{|\mathbf{k}_{1}||\mathbf{k}_{2}|}}\thinspace.\label{eq:-62}
\end{equation}
In a similar way, one can also derive the thin-slab limit of ${\cal A}_{\gamma'\to\gamma}$,
which should be identical to Eq.~\eqref{eq:-62}. One can even apply
this formalism to ${\cal A}_{\gamma\to\gamma}$ and ${\cal A}_{\gamma'\to\gamma'}$.
In classical electromagnetism, they correspond to  refraction of light
in medium, and the effect at a zero angle of incidence is simply a
phase shift, i.e., ${\cal A}_{\gamma\to\gamma}=e^{i|\tilde{\mathbf{k}}_{1}|\Delta L}$
and ${\cal A}_{\gamma'\to\gamma'}=e^{i|\tilde{\mathbf{k}}_{2}|\Delta L}$,
where $|\tilde{\mathbf{k}}_{1}|\equiv\sqrt{\omega^{2}-\Pi_{\gamma\gamma}}\approx\omega-\frac{1}{2\omega}\Pi_{\gamma\gamma}$
and $|\tilde{\mathbf{k}}_{2}|\equiv\sqrt{\omega^{2}-\Pi_{\gamma'\gamma'}}\approx\omega-\frac{1}{2\omega}\Pi_{\gamma'\gamma'}$
with $\Pi_{\gamma\gamma}$ and $\Pi_{\gamma'\gamma'}$ the self-energies
of $\gamma$ and $\gamma'$ in the medium, respectively. For simplicity,
we ignore the imaginary parts of these $\Pi$'s in this subsection.
  Therefore, when a relativistic quantum superposition of $\gamma$
and $\gamma'$  propagates through the medium, the differential conversion
amplitudes are
\begin{equation}
i\frac{d}{dL}\left[\begin{array}{cc}
{\cal A}_{\gamma\to\gamma} & {\cal A}_{\gamma'\to\gamma}\\
{\cal A}_{\gamma\to\gamma'} & {\cal A}_{\gamma'\to\gamma'}
\end{array}\right]\approx\frac{1}{2\omega}\left[\begin{array}{cc}
\Pi_{\gamma\gamma} & -\Pi_{\gamma\gamma'}\\
-\Pi_{\gamma\gamma'} & \Pi_{\gamma'\gamma'}
\end{array}\right],\label{eq:-63}
\end{equation}
where $L$ is the propagation length and we have removed the contribution
of an overall phase factor  to the diagonal elements. Eq.~\eqref{eq:-63}
can be applied to even a thick slab of medium. In this case, the last
matrix in Eq.~\eqref{eq:-63} can be viewed as an effective Hamiltonian
$H$ which governs the oscillation between $\gamma$ and $\gamma'$.
In particular, if the medium is homogeneous, the $S$ matrix of this
process reads 
\begin{equation}
e^{-iHL}=\left[\begin{array}{cc}
c_{\theta}^{2}e_{1}+s_{\theta}^{2}e_{2} & c_{\theta}s_{\theta}(e_{1}-e_{2})\\
c_{\theta}s_{\theta}(e_{1}-e_{2}) & s_{\theta}^{2}e_{1}+c_{\theta}^{2}e_{2}
\end{array}\right]\ \text{with}\ e_{1,2}\equiv e^{-\frac{iL}{2\omega}\tilde{m}_{1,2}^{2}}\thinspace.\label{eq:-65}
\end{equation}
Here $\left(s_{\theta},\ c_{\theta}\right)\equiv\left(\sin\theta,\ \cos\theta\right)$
with $\theta\equiv\frac{1}{2}\arctan\frac{2\Pi_{\gamma\gamma'}}{\Pi_{\gamma'\gamma'}-\Pi_{\gamma\gamma}}$,
and $\tilde{m}_{1,2}^{2}$ are the mass eigenvalues of $\gamma$ and
$\gamma'$ in the medium. Their exact forms are
\begin{equation}
\tilde{m}_{1,2}^{2}=\frac{\Pi_{\gamma\gamma}+\Pi_{\gamma'\gamma'}}{2}\pm\frac{1}{2}\sqrt{\left(\Pi_{\gamma\gamma}-\Pi_{\gamma'\gamma'}\right)^{2}+4\Pi_{\gamma\gamma'}^{2}}\thinspace.\label{eq:-64}
\end{equation}
If $\Pi_{\gamma\gamma'}\ll|\Pi_{\gamma\gamma}-\Pi_{\gamma'\gamma'}|$,
Eq.~\eqref{eq:-64} implies $\tilde{m}_{1}^{2}\approx\Pi_{\gamma\gamma}^{2}$
and $\tilde{m}_{2}^{2}\approx\Pi_{\gamma'\gamma'}=m_{\gamma'}^{2}+\varepsilon^{2}\Pi_{\gamma\gamma}\approx m_{\gamma'}^{2}$.
In this case, the mass squared difference is given by $\tilde{m}_{1}^{2}-\tilde{m}_{2}^{2}\approx\Pi_{\gamma\gamma}^{2}-m_{\gamma'}^{2}$.
If $\Pi_{\gamma\gamma'}\gg|\Pi_{\gamma\gamma}-\Pi_{\gamma'\gamma'}|$,
the difference becomes $\tilde{m}_{1}^{2}-\tilde{m}_{2}^{2}\approx2\Pi_{\gamma\gamma'}\approx2\varepsilon\Pi_{\gamma\gamma}$
(or equivalently, $2\varepsilon m_{\gamma'}^{2}$, as $\Pi_{\gamma\gamma}$
is close to $m_{\gamma'}^{2}$ in this case), implying that the oscillation
length near the resonance is inversely proportional to $\varepsilon$. 

The probability of $\gamma\to\gamma'$ conversion is the modulus squared
of the off-diagonal element of $e^{-iHL}$ in Eq.~\eqref{eq:-65}:
\begin{equation}
P_{\gamma\to\gamma'}=4\left|\frac{\Pi_{\gamma\gamma'}^{2}}{\tilde{m}_{1}^{2}-\tilde{m}_{2}^{2}}\right|^{2}\sin^{2}\left(\frac{\tilde{m}_{1}^{2}-\tilde{m}_{2}^{2}}{4\omega}L\right),\label{eq:-26}
\end{equation}
 which shows that the conversion probability indeed oscillates with
$L$. If $L$ is sufficiently large, using 
\begin{equation}
\delta(x)=\frac{1}{\pi}\lim_{y\to\infty}\frac{\sin^{2}(xy)}{x^{2}y}\thinspace,\label{eq:-29}
\end{equation}
we can reduce Eq.~\eqref{eq:-26} to
\begin{equation}
P_{\gamma\to\gamma'}\to\varepsilon^{2}\Pi_{\gamma\gamma}^{2}\pi\frac{L}{\omega}\delta\left(\tilde{m}_{1}^{2}-\tilde{m}_{2}^{2}\right).\label{eq:-25}
\end{equation}
Dividing it by $L$ (so that it can be interpreted as the differential
conversion probability per unit length) and multiplying it by $f_{\gamma}$
(which is related to the number of photons), we arrive at
\begin{equation}
f_{\gamma}\frac{P_{\gamma\to\gamma'}}{L}\to f_{\gamma}\varepsilon^{2}m_{\gamma'}^{4}\pi\frac{1}{\omega}\delta\left(m_{\gamma'}^{2}-\Pi_{\gamma\gamma}^{2}\right),\label{eq:-30}
\end{equation}
where we have used the approximation $\tilde{m}_{1}^{2}\approx\Pi_{\gamma\gamma}^{2}$
and $\tilde{m}_{2}^{2}\approx m_{\gamma'}^{2}$. Eq.~\eqref{eq:-30}
is essentially the same as the resonant production rate in Eq.~\eqref{eq:-17}.
This reveals the physical meaning of Eq.~\eqref{eq:-17}:  the photon
occupation number ($f_{\gamma}$) multiplied by the differential conversion
probability ($P_{\gamma\to\gamma'}/L$) gives the production rate
of the dark photon.

\section{Evolution\label{sec:Cosmological-evolution}}

The cosmological evolution of dark photons can be obtained by solving
the Boltzmann equation. In this section, we numerically solve the
Boltzmann equation for $n_{\gamma'}$ with the collision terms in
Tab.~\ref{tab:Collision-terms}. The medium effect is included using
Eq.~\eqref{eq:-16}. Note that although our discussion focuses on
the freeze-in regime, our numerical solutions obtained by solving
the Boltzmann equation incorporating both gain and loss collision
terms remain valid for completely thermalized dark photons. 

\begin{figure}
\centering

\includegraphics[width=0.99\textwidth]{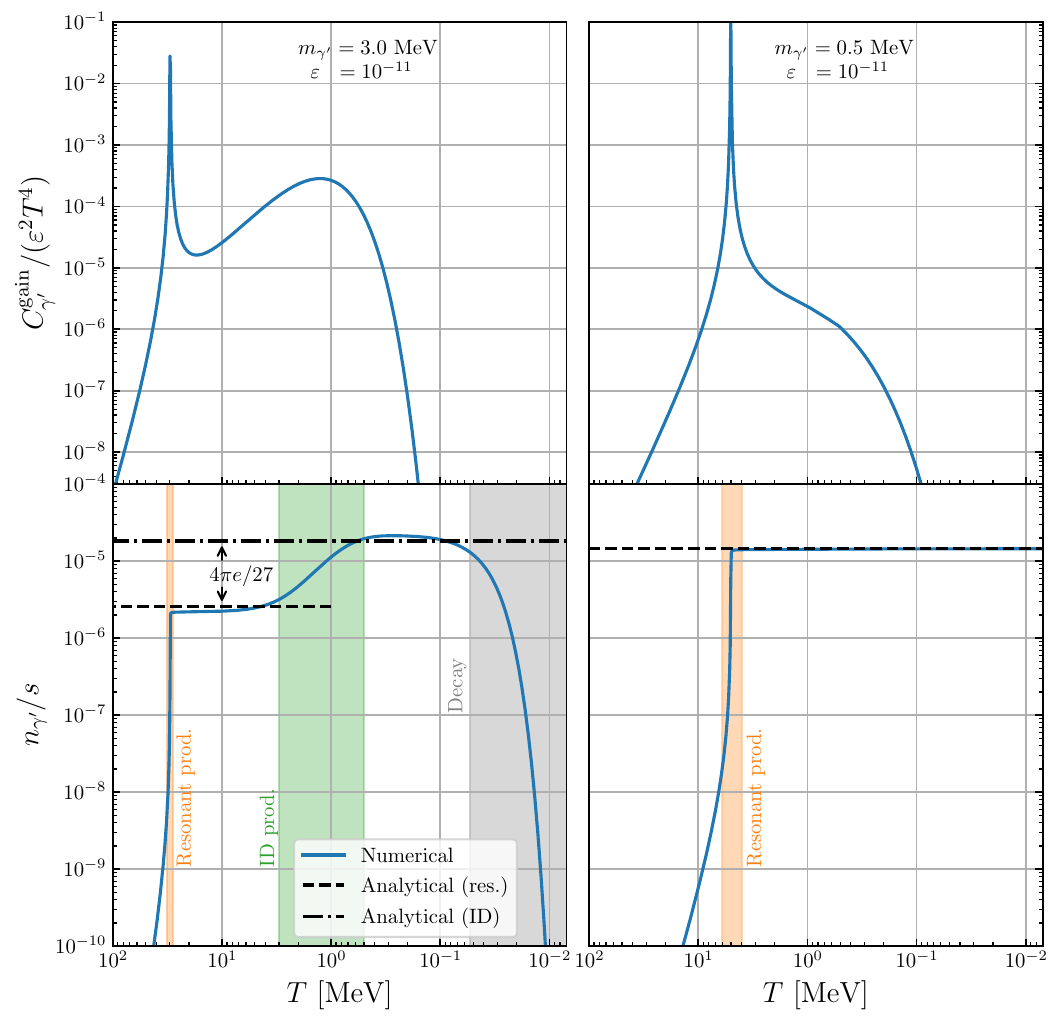}

\caption{\label{fig:sol} Upper panels: Resonances of the production rate;
Lower panels: the evolution of the dark photon number density $n_{\gamma'}$
divided by the entropy density $s$. The dashed and dash-dotted lines
represent our analytical estimates given by Eqs.~\eqref{eq:-23} and
\eqref{eq:-42}. The gap between them, $4\pi e/27\approx0.14$, is
estimated by Eq.~\eqref{eq:-59}. }
\end{figure}

In the lower panels of Fig.~\ref{fig:sol}, we show the solutions
of two benchmarks with $m_{\gamma'}\in\{3,\ 0.5\}$ MeV and $\varepsilon=10^{-11}$.
As previously discussed, when the temperature reaches a certain value,
the resonant production is achieved. Indeed, one can see that within
the orange bands shown in Fig.~\ref{fig:sol}, $n_{\gamma'}/s$ increases
rapidly. On the upper panels, we present the corresponding $C_{\gamma'}^{{\rm gain}}$-$T$
curves, which exhibit sharp peaks at $T\approx30$ and $5$ MeV. 

After the resonance, the evolution may or may not be able to produce
a significant extra amount  of dark photons, depending on whether
the inverse decay process is kinematically allowed or not. For the
benchmark with $m_{\gamma'}=3$ MeV, which is above $2m_{e}$, the
inverse decay process becomes effective during the period indicated
by the green band in Fig.\ref{fig:sol}. In this case, the post-resonance
evolution generates more dark photons through inverse decay  than
the resonance.  Hence the $n_{\gamma'}/s$ curve exhibits two plateaus,
one occurring after the resonance (after $T$ reaches about $30$
MeV) and the other occurring after the completion of inverse decay.
Both can be analytically estimated, as given by Eqs.~\eqref{eq:-23}
and \eqref{eq:-42}. The second plateau is expected to be a factor
of $4\pi e/27\approx0.14$ higher than the first, according to Eq.~\eqref{eq:-59}.
Since this benchmark is in the freeze-in regime, the dark photon is
relatively long-lived compared to the cosmic time $t\approx1/(2H)$.
After reaching the second plateau, the number of dark photons in a
comoving volume remains approximately constant until $t$ becomes
comparable to its lifetime $\tau_{\gamma'}$. At $t\gtrsim\tau_{\gamma'}$,
corresponding to the gray band in Fig.\ref{fig:sol}, the number decreases
exponentially. 

By contrast, the evolution of the benchmark with $m_{\gamma'}=0.5$
MeV (right panels in Fig.~\ref{fig:sol}) has a simpler structure:
the $n_{\gamma'}/s$ curve increases rapidly during resonant production
and reaches a stable value that can be estimated by Eq.~\eqref{eq:-23}.
The post-resonance production  makes only a negligible contribution.
Since the two-body decay channel $\gamma'\to e^{+}e^{-}$ is forbidden
for $m_{\gamma'}<2m_{e}$, the dark photon in this example is extremely
long-lived, with $\tau_{\gamma'}$ comparable to the age of universe
$\tau_{U}\approx4.35\times10^{17}\ {\rm sec}$. Consequently, the
$n_{\gamma'}/s$ curve remains essentially flat throughout the cosmic
history from the resonant production to the present epoch. The abundance
of these long-lived dark photons could be slightly reduced by $\gamma'+e^{-}\to\gamma+e^{-}$,
but this effect is negligibly small for this benchmark. More generally,
we find that as long as $\varepsilon$ lies in the freeze-in regime,
this process never becomes significant  in the evolution of dark photons.

Here we would like to comment on possible decay channels of $\gamma'$
in the low-mass regime ($m_{\gamma'}<2m_{e}$). In this regime, without
introducing extra light species, $\gamma'$ can only decay to species
lighter than electrons, namely photons and neutrinos ($\nu$). Decaying
to two photons ($\gamma'\to2\gamma$) is not allowed as this would
violate the Landau--Yang theorem.  Decaying to three photons ($\gamma'\to3\gamma$,
arising from a loop diagram) is allowed, with the following decay
width~\cite{Pospelov:2008jk}
\begin{equation}
\Gamma_{\gamma'\to3\gamma}\approx\frac{17\varepsilon^{2}\alpha^{4}}{11664000\pi^{3}}\frac{m_{\gamma'}^{9}}{m_{e}^{8}}\thinspace.\label{eq:-60}
\end{equation}
Note that Eq.~\eqref{eq:-60} is derived in the Euler-Heisenberg limit,
which implies that it may be not very accurate when $m_{\gamma'}$
approaches $2m_{e}$---see Refs.~\cite{McDermott:2017qcg,Linden:2024uph,Linden:2024fby}
for discussions on small deviations from this limit. Since a precise
calculation of $\Gamma_{\gamma'\to3\gamma}$ is not required in this
work, Eq.~\eqref{eq:-60} suffices for our analysis. Regarding the
possibility of decaying to neutrinos ($\gamma'\to2\nu$), this is
implied by Eq.~\eqref{eq:FF} where $F^{\mu\nu}$ contains a small
component of the $Z$ boson. However, the resulting coupling of $\gamma'$
to neutrinos is suppressed by $s_{W}m_{\gamma'}^{2}/m_{Z}^{2}$~\cite{Li:2023vpv,Lindner:2018kjo}.
We have estimated the decay width of $\gamma'\to2\nu$ and find that
it is sufficiently small to be neglected in this work.  

\begin{figure}
\centering

\includegraphics[width=0.49\textwidth]{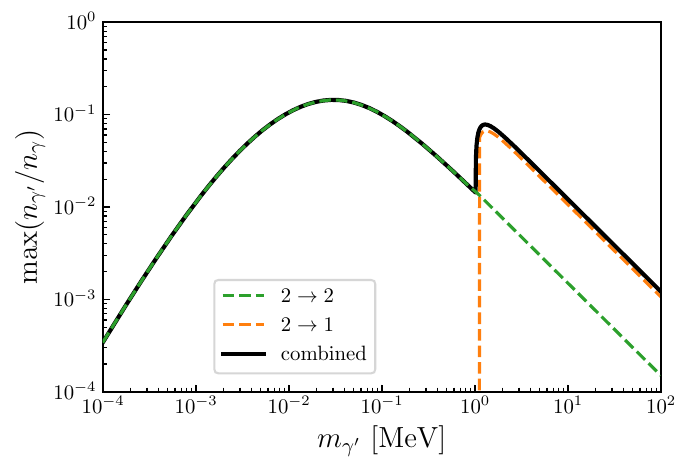}\includegraphics[width=0.49\textwidth]{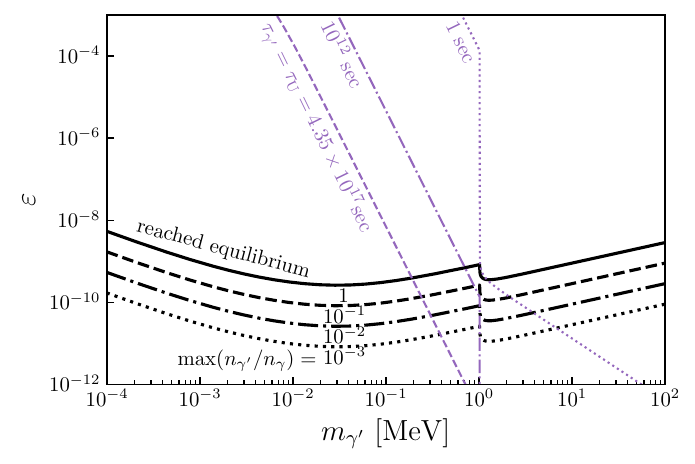}

\caption{\label{fig:yield}Left: The maximum of $n_{\gamma'}/s$ during the
evolution as a function of $m_{\gamma'}$, assuming $\varepsilon=10^{-10}$.
 Right: Contours on the $\varepsilon$-$m_{\gamma'}$ plane to indicate
whether $\gamma'$ has ever reached equilibrium (black lines) and
whether it is long-lived (purple lines). }
\end{figure}

In the left panel of  Fig.~\ref{fig:yield}, we scan over a wide
range of $m_{\gamma'}$ covering both the high-mass and low-mass regimes
with $\varepsilon=10^{-10}$ and plot $\max\left(n_{\gamma'}/n_{\gamma}\right)$
as a function of $m_{\gamma'}$.  Here ``max'' takes the maximum
of $n_{\gamma'}/n_{\gamma}$ during the evolution. This maximum is
used to quantify how close the evolution is to thermal equilibrium.
 In this plot, the curves labeled $2\to1$ and $2\to2$ represent
contributions from inverse decay and from scattering processes (semi-Compton
and annihilation), while the black curve includes both. As  shown
in this plot, the black curve is always significantly below $1$,
implying that dark photons with $\varepsilon=10^{-10}$ can never
reach thermal equilibrium. 

In the right panel of Fig.~\ref{fig:yield}, we show the contours
of $\max\left(n_{\gamma'}/n_{\gamma}\right)=10^{-3}$, $10^{-2}$,
$10^{-1}$ and $1$ on the $\varepsilon$-$m_{\gamma'}$ plane. In
the region above the black solid line, dark photons have reached thermal
equilibrium. In this plot, we also show the lifetime contours (purples
lines) corresponding to $\tau_{\gamma'}=4.35\times10^{17}\ {\rm sec}$
(the age of the Universe), $10^{12}\ \text{sec}$ (the time scale
at $T=1$ eV), and $1\ \text{sec}$ (the time scale at $T=1$ MeV).
For $\tau_{\gamma'}\lesssim1$ sec, dark photons decay after neutron
freeze-out and neutrino decoupling. For $\tau_{\gamma'}\ll10^{12}$
sec, dark photons within the shown mass range are non-relativistic
long-lived particles at recombination ($0.3$ eV) and the matter-radiation
equality ($0.8$ eV), implying that they play a role of dark matter
at this epoch. Note that within the gap between the dash-dotted ($\tau_{\gamma'}=10^{12}$
sec) and dashed ($\tau_{\gamma'}=\tau_{U}$) purple lines, dark photons
would behave as dark matter for CMB observations but cannot contribute
to the present-day dark matter

\section{Cosmological constraints\label{sec:Cosmological-constraints}}

Thermally produced dark photons  in the early Universe are subject
to cosmological constraints from BBN and CMB measurements.   Below
we discuss constraints from BBN and CMB measurements separately. 

\subsection{Constraints from BBN\label{subsec:Constraints-from-BBN}}

\begin{figure}
\centering

\includegraphics[width=0.6\textwidth]{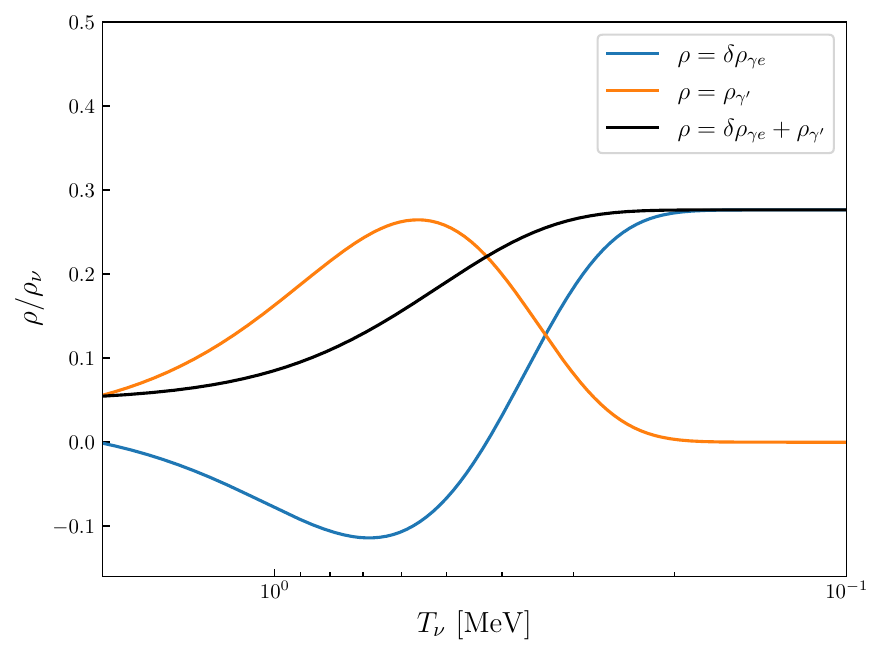}

\caption{Energy transfer between dark photons and the $\gamma$-$e^{\pm}$
plasma after neutrino decoupling, assuming $m_{\gamma'}=3\ \text{MeV}$,
$\varepsilon=2\times10^{-10}$,  and $T_{\nu\text{dec}}=2$ MeV. \label{fig:energy-injection}}
\end{figure}

BBN measurements are sensitive to new physics that alters the standard
cosmological history spanning from neutrino decoupling (roughly at
$2$ MeV) to the onset of nucleosynthesis (slightly below $0.1$ MeV).
For instance, in the presence of a new relativistic particle species
with negligible interactions (i.e. dark radiation) during this epoch,
it would contribute to the effective number of neutrino species ($N_{{\rm eff}}$),
which has been determined very precisely by BBN measurements: $N_{{\rm eff}}^{\text{BBN}}=2.898\pm0.141$~\cite{ParticleDataGroup:2024cfk}.
Therefore, any dark radiation that carries a few tens of percent of
 the neutrino energy (single flavor) could  be probed or excluded
by the present BBN observations at a considerably high confidence
level. 

When applying BBN constraints on the dark photon, it is important
to notice that its impact on BBN can be quite dynamical rather than
a static contribution to $N_{{\rm eff}}$, as its abundance may vary
significantly during the BBN epoch. In addition, it may also alter
the evolution of the $\gamma$-$e^{\pm}$ plasma via energy injection
and absorption.

 In Fig.~\ref{fig:energy-injection}, we illustrate this point with
a benchmark scenario:  $m_{\gamma'}=3\ \text{MeV}$, $\varepsilon=2\times10^{-10}$,
and the neutrino decoupling temperature $T_{\nu\text{dec}}=2$ MeV.
  Here,  $\rho_{\gamma'}$ denotes the energy density of dark photons,
and $\delta\rho_{\gamma e}$ denotes the portion being injected into
(if $\delta\rho_{\gamma e}>0$) or absorbed from (if $\delta\rho_{\gamma e}<0$)
the $\gamma$-$e^{\pm}$ plasma after neutrino decoupling. For the
convenience of $N_{{\rm eff}}$ interpretation, we plot the ratios
of them to  the neutrino energy density of a single flavor, $\rho_{\nu}$.
As is shown in Fig.~\ref{fig:energy-injection}, after neutrino decoupling,
the production of dark photons  still proceeds, until decay dominates
at sub-MeV temperatures. Since neutrinos have already decoupled, the
production and decay processes  affect only the $\gamma$-$e^{\pm}$
plasma, while the neutrino sector remains unaffected. The production
of $\gamma'$ consumes energy from the $\gamma$-$e^{\pm}$ plasma,
and the decay returns energy. As a consequence, the $\delta\rho_{\gamma e}$
curve first decreases and then increases. The final value of $\delta\rho_{\gamma e}$
is positive, implying that dark photons return more energy than they
have consumed since neutrino decoupling. The reason for this is two-fold:
(i) at neutrino decoupling, dark photons have already carried a significant
amount of energy which will eventually be released into the $\gamma$-$e^{\pm}$
plasma; (ii) the mass of dark photons leads to a significant dilution-resistant
effect~\cite{Li:2023puz}. 

The variations shown in Fig.~\ref{fig:energy-injection} suggest
that one should not straightforwardly use the BBN measurement of $N_{{\rm eff}}$
to set the corresponding bound on dark photons, and call for a dedicated
study on how the fundamental BBN predictions, i.e., the primordial
abundances of light elements, are affected by dark photons. To our
knowledge, a comprehensive study including the above non-trivial variations
and a thorough calculation of all light-element  abundances is still
lacking. In this work, we select one of the primary observables, the
helium abundance $Y_{P}$, for a quantitative study. For other elements,
the calculation can be  more involved due to photo-dissociation effects,
as we will discuss in Sec.~\ref{subsec:Combined}.

 The helium abundance $Y_{P}$ can be computed by evolving the neutron-to-baryon
ratio towards the onset of nucleosynthesis. Without involving a sophisticated
nuclear reaction network, $Y_{P}$ can be computed quite accurately
because more than $99.9\%$ neutrons present before nucleosynthesis
are ultimately incorporated into helium.  

The neutron-to-baryon ratio obeys the following Boltzmann equation:
\begin{equation}
\frac{dX_{n}}{dt}=-\Gamma_{n\to p}X_{n}+\Gamma_{p\to n}(1-X_{n})\thinspace,\label{eq:-67}
\end{equation}
where $X_{n}\equiv n_{n}/(n_{n}+n_{p})$ with $n_{n}$ and $n_{p}$
the neutron and proton number densities, respectively. The conversion
rate $\Gamma_{n\to p}$ ($\Gamma_{p\to n}$) corresponds to the probability
of a neutron (proton) being converted to a proton per unit time. Before
neutron freeze-out ($T\gtrsim1$ MeV), $X_{n}$ follows the in-equilibrium
value:
\begin{equation}
X_{n}^{({\rm eq})}\approx\frac{e^{-Q/T}}{1+e^{-Q/T}}\thinspace,\label{eq:-68}
\end{equation}
with $Q\approx1.3$ MeV the mass difference between the neutron and
the proton. We use Eq.~\eqref{eq:-68} as the initial condition, and
follow Appendix B of Ref.~\cite{Wu:2025ovd} to compute $\Gamma_{n\to p}$
and $\Gamma_{p\to n}$.  

\begin{figure}
\centering

\includegraphics[width=0.6\textwidth]{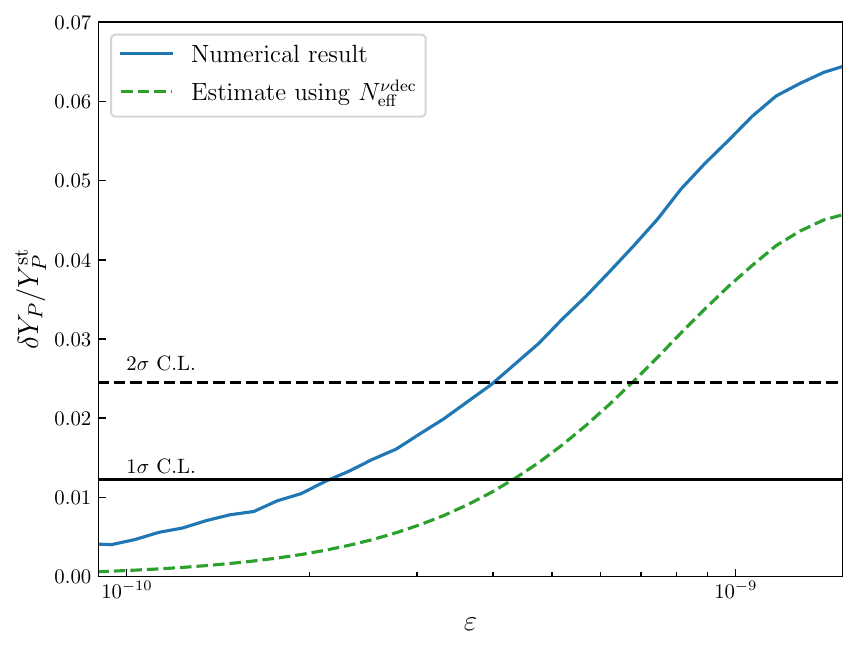}

\caption{Deviation ($\delta Y_{P}$) of the helium abundance from its standard
value $Y_{P}^{{\rm st}}$ as a function of $\varepsilon$, assuming
$m_{\gamma'}=3$ MeV. The blue line represents our numerical result
obtained by solving Eq.~\eqref{eq:-67}. The green line represents
a simple  estimate using $N_{{\rm eff}}^{\nu\text{dec}}\equiv\rho_{\gamma'}/\rho_{\nu}$
evaluated at $T=T_{\nu\text{dec}}$ and Eq.~\eqref{eq:-66}. The black
lines correspond to 1$\sigma$ and 2$\sigma$ observational bounds.
\label{fig:YP}}
\end{figure}

By solving Eq.~\eqref{eq:-67}, we obtain the value of $X_{n}$ before
nucleosynthesis and use it to compute $Y_{P}\approx2X_{n}(T_{{\rm nuc}})$
where $T_{{\rm nuc}}\approx0.08$ MeV is the temperature of nucleosynthesis.
In Fig.~\ref{fig:YP}, we present the obtained result for $m_{\gamma'}=3$
MeV (blue curve). Here the observational bounds (black lines) are
set according to the PDG recommended experimental value $Y_{P}=0.245\pm0.003$
(1$\sigma$ C.L.)~\cite{ParticleDataGroup:2024cfk}. For comparison,
we also add the green line representing a simple estimate based on
$N_{{\rm eff}}^{\nu\text{dec}}\equiv\rho_{\gamma'}/\rho_{\nu}$ evaluated
at $T=T_{\nu\text{dec}}$. If the dark photon could behave as a static
contribution to $N_{{\rm eff}}$, then its influence on $Y_{P}$ according
to Refs.~\cite{Cyburt:2015mya,Pitrou:2018cgg} could be estimated
via
\begin{equation}
Y_{P}\approx0.24703\left(\frac{N_{{\rm eff}}}{3.0}\right)^{0.163}.\label{eq:-66}
\end{equation}
As one can see from Fig.~\ref{fig:YP}, the estimate based on $N_{{\rm eff}}^{\nu\text{dec}}$
qualitatively demonstrates how $Y_{P}$ is affected by the dark photon,
but the deviation from the true value is not negligible, which is
expected from the variations shown in Fig.~\ref{fig:energy-injection}.
Therefore, $N_{{\rm eff}}$ measured by BBN should not be used to
set a quantitative bound on dark photons. 

From the blue curve, one can see that for $m_{\gamma'}=3$ MeV, $\varepsilon>4.0\times10^{-10}$
has been excluded by BBN at 2$\sigma$ C.L. Following a similar analysis
for other values of $m_{\gamma'}$, we perform a scan of the parameter
space and obtain the blue region in the left panel of Fig.~\ref{fig:result}.
This region is excluded by the BBN measurement of $Y_{P}$ at 2$\sigma$
C.L. Although the estimate based on $N_{{\rm eff}}^{\nu\text{dec}}$
is not accurate, we think it is nevertheless useful to present the
corresponding bound derived from $N_{{\rm eff}}^{\nu\text{dec}}$
due to its simplicity, as it does not require solving the neutron
evolution. This is presented in the same plot as the green dashed
contour. 

Here we would like to discuss some noteworthy features of the blue
and green contours in the left panel of Fig.~\ref{fig:result}. First,
their right edges are nearly vertical, except for a small ``nose''
in the lower-right corner. The vertical part arises from that the
energy density of fully thermalized dark photons is independent of
$\varepsilon$. The value of $\rho_{\gamma'}$ at neutrino decoupling
is Boltzmann suppressed for heavy dark photons with $m_{\gamma'}\gg T_{\nu\text{dec}}$.
So when $m_{\gamma'}$ exceeds certain values (about 7 MeV), the abundance
of such dark photons becomes negligibly small. The ``nose'' is a
consequence of the interplay between the Boltzmann suppression and
freeze-in, occurring when the system is near the boundary of thermal
equilibrium---see Fig.~2 in Ref.~\cite{Li:2023puz} for illustration.
The left edges of these contours are also very steep. The part below
the left edges corresponds to the scenario that the resonant production
occurs too late to be relevant to BBN.

\subsection{Constraints from CMB $N_{{\rm eff}}$ and DM relic abundance}

The CMB measurement of $N_{{\rm eff}}$ (denoted by $N_{{\rm eff}}^{\text{CMB}}$
in what follows), unlike the BBN measurement, can be readily applied
 to set a straightforward constraint on dark photons,  as long as
they decay before recombination. If they are sufficiently long-lived
and become non-relativistic at matter-radiation equality, which is
typical for $m_{\gamma'}<2m_{e}$, they contribute to the DM relic
abundance observed at the CMB epoch. Therefore, short- and long-lived
dark photons are subject to different constraints from CMB observations. 

For short-lived dark photons ($\tau_{\gamma'}\lesssim10^{12}$ sec),
the contribution to $N_{{\rm eff}}^{\text{CMB}}$ is negative because
the decay injects energy into the $\gamma$-$e^{\pm}$ plasma, thereby
decreasing the neutrino-to-photon temperature ratio.  To set the
constraint, we adopt the latest CMB measurement of $N_{{\rm eff}}$
from Ref.~\cite{Planck:2018vyg}: $N_{{\rm eff}}^{\text{CMB}}=2.99\pm0.17$
at $1\sigma$ C.L. Subtracting it by the standard value $3.044$~\cite{Bennett:2020zkv},
we obtain the constraint on the negative contribution: $\Delta N_{{\rm eff}}^{\text{CMB}}\gtrsim-0.4$
at $2\sigma$ C.L.  In the left panel of Fig.~\ref{fig:result},
we present the corresponding bound as the red shaded region. 

For long-lived dark photons with $\tau_{\gamma'}\gtrsim10^{12}$ sec
and $m_{\gamma'}\gtrsim1$ eV, they behave as dark matter in the CMB
epoch. The energy density of dark matter in this epoch is given by
$\rho_{{\rm DM}}\approx9.74\times10^{-12}\ \text{eV}^{4}\times(T/T_{0})^{3}$
with $T_{0}\approx2.73$ K the present CMB temperature. By requiring
$\rho_{\gamma'}/\rho_{{\rm DM}}<1$ at $T=1$ eV, we obtain the constraint
of dark matter overproduction, represented by the purple region in
the left panel of Fig.~\ref{fig:result}.

\subsection{Combined constraints  and discussions\label{subsec:Combined}}

\begin{figure}
\centering

\includegraphics[width=0.49\textwidth]{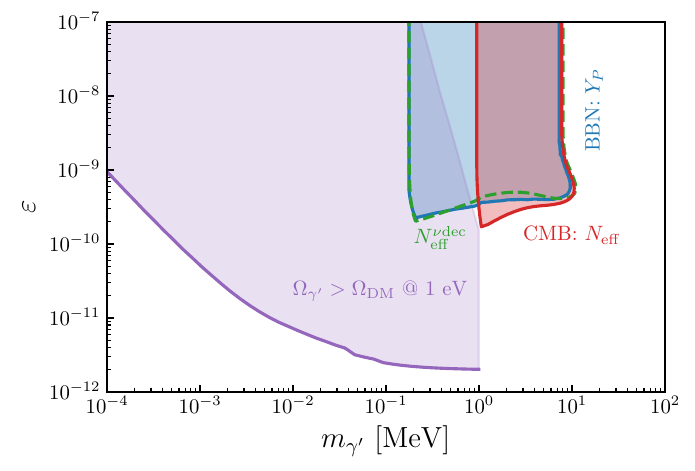}\includegraphics[width=0.49\textwidth]{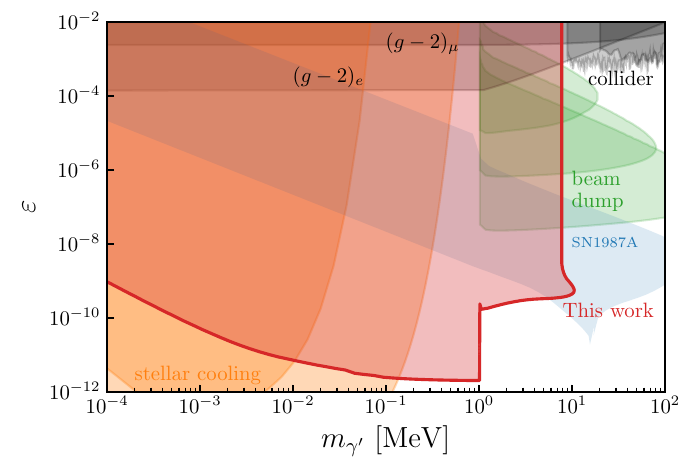}

\caption{Cosmological bounds on the dark photon derived in this work (left
panel) and compared with other known bounds (right panel). On the
left panel, we present cosmological constraints derived from the BBN
observable $Y_{P}$ (blue), $N_{{\rm eff}}$ at the CMB epoch (red),
and DM relic abundance ($\Omega_{\gamma'}>\Omega_{{\rm DM}}$ at $T=1$
eV, purple). Note that the $N_{{\rm eff}}^{\nu\text{dec}}$ contour
(green) is only a qualitative bound derived using Eq.~\eqref{eq:-66},
hence represented by a dashed line. On the right panel, we combine
the these cosmological bounds and present it as the red shaded region.
For other known bounds, see the main text for more details. \label{fig:result}
}
\end{figure}

In the right panel of Fig.~\ref{fig:result}, we combine the cosmological
bounds derived in the above analysis together and compare it with
other known bounds. Here, the orange regions are excluded by stellar
cooling, taken from Ref.~\cite{Li:2023vpv}. The blue region represents
the bound derived from supernovae, taken from Ref.~\cite{Chang:2016ntp}.
The laboratory bounds derived from $(g-2)_{\mu}$, $(g-2)_{e}$, collider,
and beam dump experiments are generated using the {\tt DARKCAST package}~\cite{Ilten:2018crw}.
 In comparison with these bounds, the combined cosmological bound
is the most stringent in the mass range from 0.1 MeV to 6 MeV. 

The cosmological bounds presented in this work are relatively conservative
but robust. While we adopt only the $^{4}\text{He}$ abundance to
set the BBN bound, one could also consider the abundances of other
light elements such as $\text{D}$, $^{3}\text{He}$, and $^{7}\text{Li}$.
In addition, energetic photons arising from dark photon decay may
cause photo-dissociation of light elements, an effect also important
for constraining unstable dark particles~\cite{Kawasaki:2017bqm,Hufnagel:2018bjp,Forestell:2018txr,Coffey:2020oir}.
 The abundances of $\text{D}$ and $^{3}\text{He}$ are more sensitive
to photo-dissociation effects than that of $^{4}\text{He}$ due to
their lower binding energies, rendering them more easily photo-dissociated.
Furthermore, due to the dominance of $^{4}\text{He}$ among the synthesized
elements, when $^{4}\text{He}$ photo-dissociation occurs, a small
fractional change in $^{4}\text{He}$ may lead to a large relative
enhancement of $\text{D}$ and $^{3}\text{He}$ while leaving the
$^{4}\text{He}$ abundance almost unchanged.  Taking $\text{D}$
for example, its abundance is four orders of magnitude lower than
that of $^{4}\text{He}$. If $0.1\%$ of $^{4}\text{He}$ is dissociated
into $\text{D}$, the abundance of $\text{D}$ would be enhanced by
one order of magnitude. The abundance of $^{4}\text{He}$, by contrast,
is almost unaffected after the $0.1\%$ dissociation, implying that
the $Y_{P}$ bound presented in Fig.~\ref{fig:result} is relatively
insensitive to photo-dissociation effects.  Including other elements
in the analysis, together with potential photo-dissociation effects,
would provide additional constraints in potentially different regions
of parameter space---see, e.g., Refs.~\cite{Fradette:2014sza,Li:2020roy}.
We leave such an analysis for future work. 

 For CMB, electromagnetic radiation injected into the background
photons may cause significant CMB spectral distortions~\cite{Hu:1993gc,Chluba:2011hw}
and anisotropy  signals~\cite{Poulin:2016anj}. We also leave dedicated
analyses of these effects to future work. 

Finally, we comment on the possibility of the baryon-to-photon ratio
$\eta_{B}$ being affected by energy injection between the BBN and
CMB epochs. As is well known, this ratio can be measured independently
from both BBN and CMB observations and the two measurements are in
excellent agreement~\cite{ParticleDataGroup:2024cfk}. Hence, this
could be used to constrain dark photons that decay between the BBN
and CMB epochs, as $\eta_{B}$ could be diluted by energy injection.
However, to generate an observable effect, the abundance of dark photons
needs to be sufficiently large, and their lifetime needs to satisfy
$1\lesssim\tau_{\gamma'}/{\rm sec}\lesssim10^{12}$. We have checked
and find no viable parameter space to allow for this possibility.

\section{Conclusion \label{sec:conclusion}}

In this work, we comprehensively studied the thermal production of
dark photons in the early universe. Three processes responsible for
the thermal production (inverse decay, annihilation, and semi-Compton)
were calculated both analytically and numerically. The analytical
expressions for the collision terms of these processes are summarized
in Tab.~\ref{tab:Collision-terms} and exhibit excellent accuracy
within their valid ranges, as shown in Fig.~\ref{fig:MC}. 

The analytical collision terms allow  us to perform various subsequent
calculations analytically. Among them, we would like to highlight
the ratio, $(n_{\gamma'}a^{3})_{{\rm res.}}/(n_{\gamma'}a^{3})_{{\rm ID}}\approx4\pi e/27\approx0.14$,
where $(n_{\gamma'}a^{3})_{{\rm res.}}$ and $(n_{\gamma'}a^{3})_{{\rm ID}}$
denote the numbers of dark photons in a comoving volume produced through
the resonance and off-resonance inverse decay, respectively. This
ratio is broadly valid and insensitive to the dark photon mass $m_{\gamma'}$
and the kinetic mixing $\varepsilon$, provided that $\varepsilon$
is in the freeze-in regime and $m_{\gamma'}>2m_{e}$.  For dark photons
with $m_{\gamma'}<2m_{e}$, we find that the off-resonance production
is negligible and the yield can be simply estimated using the resonant
production. For both light and heavy dark photons, our analytical
estimates are in  good agreement with numerical solutions of the
Boltzmann equation, as shown in Fig.~\ref{fig:sol}. 

Finally, we present the cosmological constraints on the dark photon
derived solely from energy-density considerations: they only require
that $\rho_{\gamma'}/\rho_{\nu}$ at 1 MeV and $\rho_{\gamma'}/\rho_{\text{DM}}$
at 1 eV do not exceed the limits allowed by BBN and CMB observations.
The constraints obtained in this way are relatively conservative
and robust against potential variations arising from new interactions
with other dark-sector particles. Compared with known bounds from
stellar cooling, supernovae, and laboratory searches, the cosmological
constraints are most stringent in the mass range from 0.1 MeV to 6
MeV, within which they are capable of probing kinetic mixing at the
level of $10^{-12}$--$10^{-10}$.

 We hope that the analytical estimates and cosmological constraints
presented in this work may provide useful tools for ongoing studies
of dark photons and related new particles. 

\noindent\makebox[\linewidth]{\rule{0.5\linewidth}{0.4pt}}

\noindent Note added: As we were finalizing this work, Ref.~\cite{Caputo:2025avc}
appeared on arXiv. It also focuses on the dark photon within a very
similar mass regime and contains the calculation of its thermal production
in the early universe. While many technical details in Ref.~\cite{Caputo:2025avc}
differ from ours, the main calculation procedure is similar, with
the final results in good agreement with each other. Compared to Ref.~\cite{Caputo:2025avc},
 our work is more focused on the analytical discussions. 
\begin{acknowledgments}
This work is supported in part by the National Natural Science Foundation
of China under grant No.~12141501 and also by the CAS Project for
Young Scientists in Basic Research (YSBR-099). 
\end{acknowledgments}

\appendix

\section{Phase space integrals \label{sec:two-body-integral} }

In this work, the following phase space integral is often encountered:
\begin{equation}
I_{34}=\int d\Pi_{3}d\Pi_{4}(2\pi\delta)^{4}|{\cal M}|^{2}\thinspace,\label{eq:-4}
\end{equation}
where $|{\cal M}|^{2}$ is the squared matrix element of a generic
2-to-2 process, $1+2\to3+4$. If derived from a theory respecting
Lorentz invariance, $|{\cal M}|^{2}$ should be Lorentz invariant,
and its dependence on kinematics can be fully characterized by the
three Mandelstam variables, $s=(p_{1}+p_{2})^{2}$, $t=(p_{1}-p_{3})^{2}$,
and $u=(p_{1}-p_{4})^{2}$. Then Eq.~\eqref{eq:-4} is fully Lorentz-invariant,
implying that it can be conveniently calculated in, e.g., the center-of-mass
frame, and the result remains invariant when boosted to a general
frame. 

By recasting the relevant formulae in Chapter 49: {\it Kinematics}
of Ref.~\cite{ParticleDataGroup:2024cfk} into Lorentz-invariant
forms, we obtain
\begin{equation}
\int d\Pi_{3}d\Pi_{4}(2\pi\delta)^{4}|{\cal M}|^{2}=\frac{1}{8\pi\sqrt{\lambda_{s12}}}\int_{t_{-}}^{t_{+}}|{\cal M}|^{2}dt\thinspace,\label{eq:-5}
\end{equation}
where
\begin{align}
t_{\pm} & \equiv\frac{1}{2s}\left(\kappa\pm\sqrt{\lambda_{s12}\lambda_{s34}}\right),\label{eq:-6}\\
\kappa & \equiv-s^{2}+\left(m_{1}^{2}+m_{2}^{2}+m_{3}^{2}+m_{4}^{2}\right)s-\left(m_{1}^{2}-m_{2}^{2}\right)\left(m_{3}^{2}-m_{4}^{2}\right),\label{eq:-7}\\
\lambda_{sij} & \equiv\lambda\left(s,\ m_{i}^{2},\ m_{j}^{2}\right),\label{eq:-8}
\end{align}
and $\lambda(a,\ b,\ c)\equiv a^{2}+b^{2}+c^{2}-2ab-2bc-2ca$ is the
K\"all\'en function. 

If $|{\cal M}|^{2}$ happens to be independent of $t$ and $u$, then
Eq.~\eqref{eq:-5} reduces to
\begin{equation}
\int d\Pi_{3}d\Pi_{4}(2\pi\delta)^{4}=\frac{\sqrt{\lambda_{s34}}}{8\pi s}\thinspace,\label{eq:-9}
\end{equation}
which can be used to calculate phase space integrals of $1$-to-2
or 2-to-1 processes. 

If one needs to integrate out the two initial momenta in $1+2\to3+4$
or $1+2\to3$,  then  the following integral can be used:
\begin{equation}
\int d\Pi_{1}d\Pi_{2}(2\pi\delta)^{4}|{\cal M}|^{2}=\begin{cases}
\frac{1}{8\pi\sqrt{\lambda_{s34}}}\int_{t_{-}}^{t_{+}}|{\cal M}|^{2}dt & ({\rm for}\ 1+2\to3+4)\\
\frac{\sqrt{\lambda_{s12}}}{8\pi s}|{\cal M}|^{2} & ({\rm for}\ 1+2\to3)
\end{cases}\thinspace.\label{eq:-5-1}
\end{equation}

Below,  we present an example to demonstrate the use of Eqs.~\eqref{eq:-5}
and \eqref{eq:-9}. Consider the following integral
\begin{equation}
I_{1234}=\int d\Pi_{1}d\Pi_{2}d\Pi_{3}d\Pi_{4}f_{1}f_{2}(2\pi\delta)^{4}|{\cal M}|^{2}\thinspace,\label{eq:-10}
\end{equation}
where $f_{1,2}=\exp\left(-E_{1,2}/T\right)$, $m_{1,2,3,4}=0$, and
$|{\cal M}|^{2}=(p_{1}\cdot p_{2})(p_{3}\cdot p_{4})=\frac{s^{2}}{4}$.
This integral occurs in relativistic neutrino scattering in the early
universe and its result is known: $I_{1234}=\frac{3T^{8}}{8\pi^{5}}$~\cite{Luo:2020sho,Luo:2020fdt}.
Since $|{\cal M}|^{2}$ is independent of $t$, using Eq.~\eqref{eq:-9},
we quickly obtain
\begin{align}
I_{1234} & =\int d\Pi_{1}d\Pi_{2}f_{1}f_{2}\frac{s^{2}}{4}\frac{\sqrt{\lambda_{s34}}}{8\pi s}\nonumber \\
 & =\int d\Pi_{1}d\Pi_{2}f_{1}f_{2}\frac{s^{2}}{32\pi}\nonumber \\
 & =\int\frac{4\pi p_{1}^{2}dp_{1}}{(2\pi)^{3}2p_{1}}\frac{2\pi p_{2}^{2}dp_{2}dc_{12}}{(2\pi)^{3}2p_{2}}f_{1}f_{2}\frac{4p_{1}^{2}p_{2}^{2}(1-c_{12})^{2}}{32\pi}\nonumber \\
 & =\int\frac{4\pi p_{1}^{2}dp_{1}}{(2\pi)^{3}2p_{1}}\frac{2\pi p_{2}^{2}dp_{2}}{(2\pi)^{3}2p_{2}}e^{-(p_{1}+p_{2})/T}\frac{p_{1}^{2}p_{2}^{2}}{3\pi}\nonumber \\
 & =\frac{3T^{8}}{8\pi^{5}},\label{eq:-11}
\end{align}
where $c_{12}=\cos\theta_{12}$ with $\theta_{12}$ the angle between
$\mathbf{p}_{1}$ and $\mathbf{p}_{2}$. This reproduces exactly the
expected result.

\section{Collision terms\label{sec:Collision-terms}}

The collision terms for the inverse decay process can be readily obtained
using Eq.~\eqref{eq:-5-1}, given that both $|{\cal M}|^{2}$ and
$f_{1}f_{2}=e^{-E_{1}/T}e^{-E_{2}/T}=e^{-E_{3}/T}$ in Eq.~\eqref{eq:}
factor out of the integral. Below, we present the calculations for
the annihilation and semi-Compton processes. 

\subsection{Annihilation}

The squared amplitude of annihilation $\left(e^{-}+e^{+}\rightarrow\gamma+\gamma'\right)$,
after averaging over all spins and polarizations of initial and final
states, reads
\begin{equation}
|{\cal M}_{A}|^{2}=\frac{e^{4}\varepsilon^{2}}{3}\left(\frac{t_{e}}{u_{e}}+\frac{u_{e}}{t_{e}}\right)-\frac{2e^{4}\varepsilon^{2}(2+r_{m,e})}{3}\left[\left(\frac{1}{t_{e}}+\frac{1}{u_{e}}\right)\left(\frac{1}{t_{e}}+\frac{1}{u_{e}}+1\right)-\frac{r_{m,e}}{t_{e}u_{e}}\right],\label{eq:-31}
\end{equation}
where we have used the following notations: 
\begin{align}
r_{m,e} & \equiv m_{\gamma'}^{2}/m_{e}^{2}\,,\\
s_{e} & \equiv s/m_{e}^{2}\,,\\
t_{e} & \equiv t/m_{e}^{2}-1\thinspace,\\
u_{e} & \equiv u/m_{e}^{2}-1\thinspace.
\end{align}
Here $s$, $t$, $u$ are the Mandelstam variables. 

The production rate of $\gamma'$ through this process is given by
\begin{equation}
\Gamma_{A}^{\text{gain}}=\frac{g_{1}g_{2}g_{3}}{2\omega}\int d\Pi_{1}d\Pi_{2}d\Pi_{3}\left(2\pi\delta\right)^{4}\left|\mathcal{M}_{A}\right|^{2}f_{1}f_{2}\,,\label{eq:-32}
\end{equation}
where $f_{1}f_{2}=e^{-\left(E_{1}+E_{2}\right)/T}=e^{-\left(E_{3}+\omega\right)/T}$
allows us to pull it out of the integral $\int d\Pi_{1}d\Pi_{2}$.

Using Eq.~\eqref{eq:-5-1}, we first integrate out $d\Pi_{1}$ and
$d\Pi_{2}$:
\begin{equation}
\begin{alignedat}{1}I_{12}= & \int d\Pi_{1}d\Pi_{2}\left(2\pi\delta\right)^{4}\left|\mathcal{M}_{A}\right|^{2}\\
= & \frac{\sqrt{\lambda_{s12}}}{32\pi^{2}s}\int d\Omega\left|\mathcal{M}_{A}\right|^{2}\\
= & \frac{1}{32\pi^{2}}\sqrt{1-\frac{4m_{e}^{2}}{s}}\int d\Omega\left|\mathcal{M}_{A}\right|^{2}\\
= & \frac{2\left(r_{m,e}^{2}-4r_{m,e}+4\text{\ensuremath{s_{e}}}+s_{e}^{2}-8\right)\tanh^{-1}z_{e}-\left(r_{m,e}^{2}+4\text{\ensuremath{s_{e}}}+s_{e}^{2}\right)z_{e}}{12\pi(\text{\ensuremath{r_{m,e}}}-\text{\ensuremath{s_{e}}})^{2}/\left(e^{4}\varepsilon^{2}\right)}\,,
\end{alignedat}
\end{equation}
with $z_{e}\equiv\sqrt{1-4/s_{e}}$. 

Then Eq.~\eqref{eq:-32} reduces to
\begin{equation}
\begin{alignedat}{1}\Gamma_{A}^{\text{gain}}= & \frac{g_{1}g_{2}g_{3}}{2\omega}e^{-\omega/T}\int d\Pi_{3}\,e^{-E_{3}/T}I_{12}\\
= & \frac{g_{1}g_{2}g_{3}}{2\omega}e^{-\omega/T}\times\frac{1}{8\pi^{2}}\int p_{3}dp_{3}d\cos\theta\,e^{-E_{3}/T}I_{12}\,,
\end{alignedat}
\end{equation}
where $\theta$ is the angel between $\boldsymbol{p}_{4}$ and $\boldsymbol{p}_{3}$. 

Next, we change the integration variables using 
\begin{equation}
p_{3}dp_{3}d\cos\theta=\frac{ds\,dp_{3}}{2p_{4}}\,,\label{eq:-38}
\end{equation}
which can be obtained from 
\begin{equation}
s=m_{\gamma'}^{2}+2p_{3}\left(\omega-p_{4}\cos\theta\right).\label{eq:s-p-cos}
\end{equation}
 Note that the kinematics of this process should always satisfy
\begin{equation}
s\geq4m_{e}^{2}\,.
\end{equation}
Given fixed values of $\omega$ and $s$, $p_{3}$ cannot take arbitrary
values, otherwise $\cos\theta$ calculated from Eq.~\eqref{eq:s-p-cos}
may reach unphysical values.  From Eq.~\eqref{eq:s-p-cos}, one can
derive the integration interval of $p_{3}$ 
\begin{equation}
p_{3}\in\frac{1}{2}\left[\frac{s-m_{\gamma'}^{2}}{\omega+p_{4}},\frac{s-m_{\gamma'}^{2}}{\omega-p_{4}}\right]\equiv\left[p_{L},\ p_{H}\right].
\end{equation}

With the integration interval determined, we can integrate out $p_{3}$:
\begin{align}
\Gamma_{A}^{\text{gain}} & =\frac{g_{1}g_{2}g_{3}e^{-\omega/T}}{32\pi^{2}\omega p_{4}}\int_{4m_{e}^{2}}^{\infty}ds\int_{p_{L}}^{p_{H}}dp_{3}e^{-p_{3}/T}I_{12}\nonumber \\
 & =\frac{g_{1}g_{2}g_{3}Te^{-\omega/T}}{32\pi^{2}\omega p_{4}}\int_{4m_{e}^{2}}^{\infty}ds\,\left(e^{-p_{L}/T}-e^{-p_{H}/T}\right)I_{12}\,.\label{eq:-34}
\end{align}

\noindent $\blacksquare$ High-temperature limit:

When $T\gg m_{e}$,  we approximate $\left|\mathcal{M}_{A}\right|^{2}$
and $I_{12}$ as 
\begin{equation}
\left|\mathcal{M}_{A}\right|^{2}\simeq\frac{16\pi^{2}\alpha^{2}\varepsilon^{2}}{3}\left(\frac{t}{u}+\frac{u}{t}\right),
\end{equation}
\begin{equation}
I_{12}\simeq\frac{4\pi\alpha^{2}\varepsilon^{2}}{3}\left[\log\left(s/m_{e}^{2}\right)-1\right].\label{eq:-33}
\end{equation}
Substituting Eq.~\eqref{eq:-33} into Eq.~\eqref{eq:-34}, we obtain
\begin{equation}
\Gamma_{A}^{\text{gain}}\simeq-g_{1}g_{2}g_{3}\frac{\alpha^{2}\varepsilon^{2}T^{2}}{6\pi\omega p_{4}}\left[p_{4}\left(1+\gamma_{E}-\log T_{e}\right)-\frac{\omega_{H}}{2}\log\frac{2\omega_{H}}{m_{e}}+\frac{\omega_{L}}{2}\log\frac{2\omega_{L}}{m_{e}}\right]e^{-\frac{\omega}{T}}\thinspace,\label{eq:-35}
\end{equation}
where 
\begin{equation}
\omega_{H}=\omega+p_{4}\,,\qquad\omega_{L}=\omega-p_{4}\,.
\end{equation}
If we neglect $m_{\gamma'}$ such that $\omega_{H}=2\omega$ and $\omega_{L}=0$,
Eq.~\eqref{eq:-35} reduces to
\begin{equation}
\Gamma_{A}^{\text{gain}}\simeq g_{1}g_{2}g_{3}\frac{\alpha^{2}\varepsilon^{2}T^{2}e^{-\omega/T}}{6\pi\omega}\left[\log\left(\frac{4\omega T}{m_{e}^{2}}\right)-\gamma_{E}-1\right].
\end{equation}
Further integrating out the phase space of $\gamma'$, we obtain
\begin{equation}
C_{A}^{\text{gain}}\simeq g_{1}g_{2}g_{3}g_{\gamma'}\frac{\alpha^{2}\varepsilon^{2}T^{4}}{6\pi^{3}}\left[-\gamma_{E}+\log\left(2T_{e}\right)\right].
\end{equation}
\noindent $\blacksquare$ Low-temperature limit:

At low temperatures ($T\ll m_{e}$), $I_{12}$ is approximated as
\begin{equation}
I_{12}\simeq\frac{2\pi\alpha^{2}\varepsilon^{2}}{3}\sqrt{s_{e}-4}\,.
\end{equation}
Substituting it into Eq.~\eqref{eq:-34}, we obtain
\begin{equation}
\Gamma_{A}^{\text{gain}}\simeq g_{1}g_{2}g_{3}\frac{\alpha^{2}\varepsilon^{2}T^{5/2}\omega_{H}^{3/2}}{24\sqrt{2}\pi^{1/2}m_{e}\omega p_{4}}e^{-\omega/T}\exp\left[-\frac{4m_{e}^{2}-m_{\gamma'}^{2}}{2\omega_{H}T}\right].
\end{equation}
Neglecting the dark photon mass, this reduces to 
\begin{equation}
\Gamma_{A}^{\text{gain}}\simeq g_{1}g_{2}g_{3}\frac{\alpha^{2}\varepsilon^{2}T^{5/2}}{12\pi^{1/2}m_{e}\omega^{1/2}}e^{-\frac{\omega^{2}+m_{e}^{2}}{\omega T}}\,.
\end{equation}
The corresponding low-$T$ limit of $C_{A}^{\text{gain}}$ becomes
\begin{equation}
C_{A}^{\text{gain}}\simeq g_{1}g_{2}g_{3}g_{\gamma'}\frac{\alpha^{2}\varepsilon^{2}m_{e}^{4}\left(4T_{e}^{3}+6T_{e}^{4}+3T_{e}^{5}\right)}{96\pi^{2}}e^{-2m_{e}/T}\,.
\end{equation}

\subsection{Semi-Compton}

The squared amplitude of semi-Compton $\left(e^{-}+\gamma\rightarrow e^{-}+\gamma'\right)$
can be easily obtained using crossing symmetry:
\begin{equation}
\left|{\cal M}_{S}\right|^{2}=-\left|{\cal M}_{A}\right|^{2}\left(t\leftrightarrow s\right)\,,
\end{equation}
where the subscripts ``$S$'' and ``$A$'' indicate semi-Compton
and annihilation processes. 

The production rate of $\gamma'$ through this process is given by
\begin{equation}
\Gamma_{S}^{\text{gain}}\equiv\frac{g_{1}g_{2}g_{3}}{2\omega}\int d\Pi_{1}d\Pi_{2}d\Pi_{3}\left(2\pi\delta\right)^{4}\left|{\cal M}_{S}\right|^{2}f_{1}f_{2}\,.
\end{equation}
Since this process may be important when the electron asymmetry becomes
significant, we keep the chemical potential of electrons in the calculation.
Similar to the calculation for the annihilation process, we first
 write the integral as
\begin{equation}
\Gamma_{S}^{\text{gain}}=g_{1}g_{2}g_{3}\frac{e^{\mu_{e^{-}}/T}e^{-\omega/T}}{2\omega}\int d\Pi_{3}e^{-E_{3}/T}I_{12}\,,\label{eq:-37}
\end{equation}
with 
\begin{equation}
I_{12}=\frac{1}{8\pi\sqrt{\lambda_{s34}}}\int dt\left|{\cal M}_{S}\right|^{2}\thinspace.\label{eq:-36}
\end{equation}
Due to 
\begin{equation}
d\Pi_{3}\rightarrow\frac{p_{3}ds\,dp_{3}}{16\pi^{2}p_{4}E_{3}}=\frac{ds\,dE_{3}}{16\pi^{2}p_{4}}\,,
\end{equation}
we rewrite Eq.~\eqref{eq:-37} as 
\begin{equation}
\Gamma_{S}^{\text{gain}}=g_{1}g_{2}g_{3}\frac{e^{\mu_{e^{-}}/T}e^{-\omega/T}}{32\pi^{2}\omega p_{4}}\int dE_{3}dse^{-E_{3}/T}I_{12}\,.\label{eq:-39}
\end{equation}

\noindent $\blacksquare$ High-temperature limit:

At high temperatures ($T\gg m_{e}$), $I_{12}$ can be approximated
as
\begin{equation}
I_{12}\simeq\frac{\pi\alpha^{2}\varepsilon^{2}}{3}\left[1+2\log\left(\frac{s}{m_{e}^{2}}\right)\right].
\end{equation}
Since this is a relatively simple function of $s$, we choose to first
integrate out $s$ and then integrate out $E_{3}$ in Eq.~\eqref{eq:-39}.
The integration interval of $s$, when $E_{3}$ is fixed as a given
value, is given by
\begin{equation}
s\in m_{e}^{2}+m_{\gamma'}^{2}+2E_{3}\omega+2p_{3}p_{4}\left[-1,1\right].
\end{equation}
After integrating out $s$, we integrate $p_{3}$ from $0$ to $\infty$
 and obtain
\begin{equation}
\Gamma_{S}^{\text{gain}}\simeq g_{1}g_{2}g_{3}\frac{\alpha^{2}\varepsilon^{2}T^{2}}{24\pi\omega}e^{\mu_{e^{-}}/T}e^{-\omega/T}\left[2\log\left(4T_{e}\frac{\omega}{m_{e}}\right)-2\gamma_{E}+1\right].
\end{equation}
Further integrating out the phase space of $\gamma'$, we obtain
\begin{equation}
C_{S}^{\text{gain}}\simeq g_{1}g_{2}g_{3}g_{\gamma'}\frac{\alpha^{2}\varepsilon^{2}T^{4}}{48\pi^{3}}\left[3-4\gamma_{E}+4\log\left(2T_{e}\right)\right].
\end{equation}

\noindent $\blacksquare$ Low-temperature limit:

At low temperatures when electrons becomes non-relativistic, $I_{12}$
can be approximated as
\begin{equation}
I_{12}\simeq\frac{\pi\alpha^{2}\varepsilon^{2}}{3\left(s_{e}-1\right)^{2}s_{e}^{2}}\left(-1+2s_{e}-16s_{e}^{2}+14s_{e}^{3}+s_{e}^{4}+\left(-6s_{e}^{2}-12s_{e}^{3}+2s_{e}^{4}\right)\log s_{e}\right),\label{eq:-40}
\end{equation}
where we have neglected the dark photon mass. 

Different from the calculation in the high-temperature limit, here
we choose to first integrate out $E_{3}$ and then integrate out $s$
in Eq.~\eqref{eq:-39}. The integration intervals of $E_{3}$ and
$s$ in this integration order are given by 
\begin{equation}
E_{3}\in\left[E_{3}^{{\rm min}},\infty\right)\ \ {\rm with}\ \ E_{3}^{{\rm min}}\equiv\frac{\omega}{s_{e}-1}+m_{e}^{2}\frac{s_{e}-1}{4\omega}
\end{equation}
and 
\begin{equation}
s\in\left(m_{e}^{2},\infty\right).
\end{equation}

First, let us integrate out $E_{3}$: 
\begin{equation}
\begin{alignedat}{1}\Gamma_{S}^{\text{gain}}\simeq & g_{1}g_{2}g_{3}\frac{m_{e}e^{\mu_{e^{-}}/T}e^{-\omega/T}}{32\pi^{2}\omega_{e}^{2}}\int ds_{e}\,I_{12}\,\int dE_{3}\frac{1}{m_{e}}e^{-E_{3}/T}\\
= & g_{1}g_{2}g_{3}\frac{Te^{\mu_{e^{-}}/T}e^{-\omega/T}}{32\pi^{2}\omega_{e}^{2}}\int_{1}^{\infty}ds_{e}\,I_{12}\,e^{-E_{3}^{{\rm min}}/T}\,.
\end{alignedat}
\end{equation}

Next, we integrate out $s$ using the non-relativistic approximation
for the final state electron:
\begin{equation}
E_{3}^{{\rm min}}\simeq m_{e}+m_{e}^{3}\frac{\left(s_{e}-1-2\omega/m_{e}\right)^{2}}{8\omega^{2}}\,.
\end{equation}
This leads to
\begin{equation}
\Gamma_{S}^{\text{gain}}\simeq g_{1}g_{2}g_{3}e^{\mu_{e^{-}}/T}e^{-\frac{\omega+m_{e}}{T}}\frac{\sqrt{2}\alpha^{2}\varepsilon^{2}T^{3/2}}{9\sqrt{m_{e}}\pi^{1/2}}\,.
\end{equation}
Then it is straightforward to obtain the low-$T$ limit of $C_{S}^{\text{gain}}$
:
\begin{equation}
C_{S}^{\text{gain}}\simeq g_{1}g_{2}g_{3}g_{\gamma'}e^{\mu_{e^{-}}/T}e^{-m_{e}/T}\frac{\sqrt{2}\alpha^{2}\varepsilon^{2}T^{9/2}}{9\sqrt{m_{e}}\pi^{5/2}}\,.
\end{equation}

\section{The photon self-energy in the thermal bath\label{sec:Pi-gg}}

In this appendix, we present the calculation of the photon self-energy
in the thermal bath of the early Universe during the epoch of $1\ \text{eV}\lesssim T\lesssim100\ \text{MeV}$.
As the temperature decreases, the thermal plasma evolves from the
relativistic to  non-relativistic regime, and the electron asymmetry
eventually becomes important. The most general formalism to evaluate
the photon self-energy, to the leading order in $\alpha$, reads~\cite{Braaten:1993jw}:
\begin{equation}
\Pi^{\mu\nu}=16\pi\alpha\int\frac{d^{3}p}{(2\pi)^{3}2E_{p}}\left[f_{e^{+}}(p)+f_{e^{-}}(p)\right]\frac{N^{\mu\nu}}{(p\cdot k)^{2}-\frac{1}{4}(k\cdot k)^{2}}\thinspace,\label{eq:-43}
\end{equation}
with
\begin{equation}
N^{\mu\nu}\equiv\left(p\cdot k\right)\left(p^{\mu}k^{\nu}+p^{\mu}k^{\nu}\right)-(k\cdot k)p^{\mu}p^{\nu}-g^{\mu\nu}\left(p\cdot k\right)^{2},\label{eq:-44}
\end{equation}
where $k$ and $p$ denote the photon and electron momenta, respectively.
Note that $p$ is an on-shell momentum while $k$ can be off-shell.
The phase space distribution functions $f_{e^{\pm}}(p)$ is given
by
\begin{equation}
f_{e^{\pm}}(p)=\frac{1}{e^{(E_{p}-\mu_{e^{\pm}})/T}+1}\thinspace.\label{eq:-45}
\end{equation}
Note that $\Pi^{\mu\nu}k_{\mu}\propto N^{\mu\nu}k_{\mu}=0$, allowing
us to decompose $\Pi^{\mu\nu}$ as follows
\begin{equation}
\Pi^{\mu\nu}=\Pi_{L}\epsilon_{L}^{\mu}\epsilon_{L}^{\nu}+\Pi_{T}\left(\epsilon_{T1}^{\mu}\epsilon_{T1}^{\nu}+\epsilon_{T2}^{\mu}\epsilon_{T2}^{\nu}\right),\label{eq:-46}
\end{equation}
where $\epsilon_{L}$ and $\epsilon_{T1,2}$ are longitudinal and
transverse polarization vectors, satisfying $\epsilon^{\mu}\epsilon_{\mu}=-1$
and $k^{\mu}\epsilon_{\mu}=0$. Here and in what follows, we omit
the subscripts ``$L$'' and ``$T1,2$'' in identities that apply
universally to all the polarization vectors. 

In order to extract the $\Pi_{L}$ and $\Pi_{T}$ components from
$\Pi^{\mu\nu}$, we use $\epsilon^{\mu}\epsilon^{\nu}$ as a projector
since the three polarization vectors are orthonormal, which implies
\begin{equation}
\Pi_{L}=\Pi_{\mu\nu}\epsilon_{L}^{\mu}\epsilon_{L}^{\nu}\thinspace,\ \ \Pi_{T}=\Pi_{\mu\nu}\epsilon_{T1}^{\mu}\epsilon_{T1}^{\nu}\thinspace.\label{eq:-47}
\end{equation}
Multiplying $N^{\mu\nu}$ with the projector, we obtain
\begin{equation}
N_{\mu\nu}\epsilon^{\mu}\epsilon^{\nu}=(p\cdot k)^{2}-(k\cdot k)\left(p\cdot\epsilon\right){}^{2}.\label{eq:-48}
\end{equation}
Without loss of generality, one can assume that $k$ aligns with the
$z$-axis and express all relevant four-vectors as follows: 
\begin{align}
k^{\mu} & =(\omega,\ 0,\ 0,\ k)\thinspace,\label{eq:-49}\\
p^{\mu} & =(E_{p},\ ps_{\theta}c_{\phi},\ ps_{\theta}s_{\phi},\ pc_{\theta})\thinspace,\label{eq:-50}
\end{align}
\begin{equation}
\epsilon_{T1}^{\mu}=(0,1,0,0)\thinspace,\ \epsilon_{T2}^{\mu}=(0,0,1,0)\thinspace,\ \epsilon_{L}^{\mu}=\frac{1}{\sqrt{\omega^{2}-k^{2}}}(k,0,0,\omega)\thinspace,\label{eq:-51}
\end{equation}
with $(s_{\theta},\ c_{\theta})\equiv\left(\sin\theta,\ \cos\theta\right)$
and $(s_{\phi},\ c_{\phi})\equiv\left(\sin\phi,\ \cos\phi\right)$. 

Using the explicit forms in Eqs.~\eqref{eq:-49}-\eqref{eq:-51},
we get
\begin{align*}
p\cdot k & =E_{p}\omega-c_{\theta}kp\thinspace,\\
p\cdot\epsilon_{L} & =\frac{E_{p}k-c_{\theta}p\omega}{\sqrt{\omega^{2}-k^{2}}}\thinspace,\\
p\cdot\epsilon_{T1} & =-pc_{\phi}s_{\theta}\thinspace.
\end{align*}
So Eq.~\eqref{eq:-48} gives
\begin{align}
N_{\mu\nu}\epsilon_{L}^{\mu}\epsilon_{L}^{\nu} & =\left(\omega^{2}-k^{2}\right)\left(E_{p}^{2}-c_{\theta}^{2}p^{2}\right),\label{eq:-52}\\
N_{\mu\nu}\epsilon_{T1}^{\mu}\epsilon_{T1}^{\nu} & =\left(c_{\theta}kp-E_{p}\omega\right)^{2}-c_{\phi}^{2}s_{\theta}^{2}p^{2}\left(\omega^{2}-k^{2}\right).\label{eq:-53}
\end{align}

Combining Eq.~\eqref{eq:-52} with Eq.~\eqref{eq:-43} and neglecting
the factor $(k\cdot k)^{2}$ in the denominator of Eq.~\eqref{eq:-43}
(see Appendix~A.2 in \cite{Braaten:1993jw}), we obtain
\begin{align}
\Pi_{L} & =\frac{\alpha}{\pi^{2}}\int\frac{2\pi p^{2}dpdc_{\theta}}{E_{p}}\left(f_{e^{+}}+f_{e^{-}}\right)\frac{\left(\omega^{2}-k^{2}\right)\left(E_{p}^{2}-c_{\theta}^{2}p^{2}\right)}{\left(E_{p}\omega-c_{\theta}kp\right)^{2}}\nonumber \\
 & =\frac{4\alpha}{\pi}\int\frac{p^{2}dp}{E_{p}}\left(f_{e^{+}}+f_{e^{-}}\right)\frac{\omega^{2}-k^{2}}{k^{2}}\left[\frac{\omega}{kv}\log\left(\frac{\omega+kv}{\omega-kv}\right)-\frac{\omega^{2}-k^{2}}{\omega^{2}-k^{2}v^{2}}-1\right],\label{eq:-54}
\end{align}
where $v\equiv p/E_{p}$.

Combing Eq.~\eqref{eq:-53} with Eq.~\eqref{eq:-43}, we obtain
\begin{align}
\Pi_{T} & =\frac{\alpha}{\pi^{2}}\int\frac{p^{2}dpd\Omega}{E_{p}}\left(f_{e^{+}}+f_{e^{-}}\right)\frac{\left(c_{\theta}kp-E_{p}\omega\right)^{2}-c_{\phi}^{2}s_{\theta}^{2}p^{2}\left(\omega^{2}-k^{2}\right)}{\left(E_{p}\omega-c_{\theta}kp\right)^{2}}\nonumber \\
 & =\frac{4\alpha}{\pi}\int\frac{p^{2}dp}{E_{p}}\left(f_{e^{+}}+f_{e^{-}}\right)\left[\frac{\omega^{2}}{k^{2}}-\frac{\omega^{2}-k^{2}}{k^{2}}\cdot\frac{\omega}{2kv}\log\left(\frac{\omega+kv}{\omega-kv}\right)\right].\label{eq:-55}
\end{align}

Eqs.~\eqref{eq:-54} and ~\eqref{eq:-55} are ready for straightforward
numerical evaluation. Nevertheless, it is worth mentioning the relativistic
and nonrelativistic limits of $\Pi_{L}$ and $\Pi_{T}$. 

When $f_{e^{\pm}}$ are nonrelativistic, we replace $\int p^{2}dpf_{e^{\pm}}\to\pi^{2}n_{e^{\pm}}$
and take $v\to0$ in Eqs.~\eqref{eq:-54} and ~\eqref{eq:-55}. The
resulting nonrelativistic limits are
\begin{align}
\Pi_{T} & \approx4\pi\alpha\frac{n_{e^{+}}+n_{e^{-}}}{m_{e}}\thinspace,\label{eq:-56}\\
\Pi_{L} & \approx4\pi\alpha\frac{n_{e^{+}}+n_{e^{-}}}{m_{e}}\left(1-\frac{k^{2}}{\omega^{2}}\right).
\end{align}

When $f_{e^{\pm}}$ are relativistic, we neglect the electron mass
(hence $v\to1$) and the chemical potential. This allows us to replace
$\int\frac{p^{2}}{E_{p}}dpf_{e^{\pm}}\to\pi^{2}T^{2}/6$.   The
resulting relativistic limits are
\begin{align}
\Pi_{T} & \approx\frac{2\pi\alpha T^{2}}{3}\left(\frac{\omega^{2}}{k^{2}}-\frac{\omega}{2k}\frac{\omega^{2}-k^{2}}{k^{2}}\log\left(\frac{\omega+k}{\omega-k}\right)\right),\label{eq:-56-1}\\
\Pi_{L} & \approx\frac{2}{3}\pi\alpha T^{2}\left(\frac{\omega^{2}}{k^{2}}-1\right)\left(\frac{\omega}{k}\log\left(\frac{\omega+k}{\omega-k}\right)-2\right).\label{eq:-56-2}
\end{align}
For $k\ll\omega$, they reduce to
\begin{equation}
\Pi_{T}\approx\Pi_{L}\approx\frac{4\pi\alpha}{9}T^{2}\thinspace.\label{eq:-57}
\end{equation}

\begin{figure}
\centering

\includegraphics[width=0.6\textwidth]{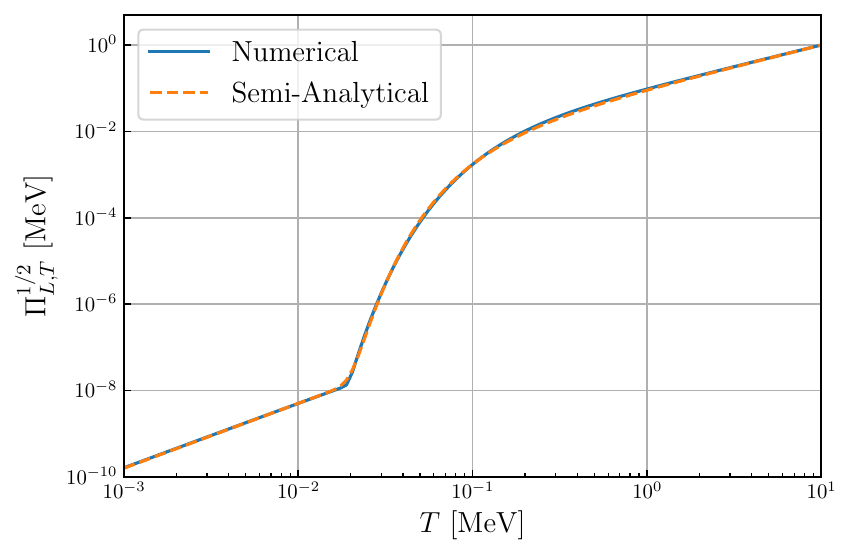}\caption{\label{fig:plasmon-mass} Numerical evaluation of $\Pi_{T,L}$ compared
with the our semi-analytical expression given by Eqs.~\eqref{eq:RePi-ana}
and \eqref{eq:-58}. }
\end{figure}

Therefore it is convenient to parametrized $\Pi_{T,L}$ as follows:
\begin{equation}
\Pi_{T,L}=\frac{4\pi\alpha}{9}T^{2}F_{T,L}^{2},\label{eq:RePi-ana}
\end{equation}
where $F_{T,L}$ accounts for the deviation of $\Pi_{T,L}$ from their
relativistic limit. By integrating Eqs.~\eqref{eq:-54} and \eqref{eq:-55}
numerically, we can obtain $\Pi_{T,L}$ and extract $F_{T,L}$ from
it. In practice, we find that the following expression can fit the
numerical values very accurately:
\begin{equation}
F_{T,L}\approx5\times10^{-5}\sqrt{x}+\exp\left(-\frac{x^{-6/5}}{9}\right),\label{eq:-58}
\end{equation}
with $x\equiv T/\text{MeV}$. Figure~\ref{fig:plasmon-mass} shows
the accuracy of the above expression in comparison with the numerical
results. 

\bibliographystyle{JHEP}
\bibliography{ref}

\end{document}